\documentclass[preprint,nofootinbib,aps,superscriptaddress,floatfix]{revtex4}
\usepackage{amsmath}
\usepackage[dvips]{graphics,color}
\begin{document}

\def\LQCD{\Lambda_{\rm QCD}}
\def\lqcd{\Lambda_{\rm QCD}}
\def\xslash#1{{\rlap{$#1$}/}}
\newcommand{\orderalpha}{ {\cal{O}}(\alpha_s)}
\newcommand{\orderalphasqr}{ {\cal{O}}(\alpha_s^2)}
\newcommand{\btoc}{\bar{B} \to \rm X_c \, \ell \, \bar{\nu}}
\newcommand{\nn}{\nonumber}


\title{A Higgs-Higgs bound state due to New Physics at a {\rm TeV}.}

\author{Benjamin Grinstein and Michael Trott}

\affiliation{Department of Physics, University of California at San Diego,\\[-5pt]
  La Jolla, CA, 92093}

\preprint{UCSD/PTH 07-03}

\begin{abstract}
We examine the effects of new physics on the Higgs sector of the standard model,
focusing on the effects on the Higgs self couplings.
We demonstrate that a low mass higgs, $m_h < 2 \,  m_t$,  can have a strong effective self coupling 
due to the effects of a new interaction at a $\rm TeV$. We investigate the possibility that 
the first evidence of such an interaction could be a higgs-higgs bound state.
To this end, we construct an effective field theory formalism to examine the physics of such a low mass higgs boson.
We explore the possibility of a non relativistic bound state of the higgs field (Higgsium) at LHC and construct a non 
relativistic effective field theory of the higgs sector that is appropriate for such studies (NRHET).

\end{abstract}
\maketitle

\section{Introduction}\label{HET}

Currently, global fits to all precision electro weak give the higgs
mass to be 
$113\genfrac{}{}{0pt}{1}{+56}{-40} {\rm GeV} $ with an upper bound given by $
m_h < 241 \, {\rm GeV}$ at $95 \% $ CL (see, {\it e.g.},  J. Erler and P. Langacker
in Sec.~10 of Ref.~\cite{Eidelman:2004wy}). LEP has also placed a lower
bound limit of $ m_h>114.4 \, {\rm GeV}$ \cite{Barate:2003sz}. 
Assuming the Standard Model (SM) of electroweak
interactions, one expects  that the higgs will soon be found at LHC.

However, there are at least two reasons why the SM with a single higgs
doublet is expected to be incomplete. The first is the triviality
problem. This asserts that the higgs self interaction, and hence its
mass, must vanish unless the theory has a finite cut-off. 
Triviality has been rigorously established only for simpler models,
but it is widely believed to hold for the SM higgs. The other is the
hierarchy problem that quadratic divergences need to be finely tuned
to keep the scale of electroweak breaking smaller than the natural
cut-off of the theory (which in the absence of new physics would be
the Planck scale).

In this paper, we investigate the effects that new physics, invoked to cure these
problems, may have on the higgs sector of the SM. We assume that the
scale of the masses of new quanta, $\mathcal{M}$, is sufficiently
higher than the scale of electro-weak symmetry breaking ($v \sim 246 \,
{\rm GeV}$) so that the quanta of the unknown new physics can be
integrated out.  
 As we want
this new physics to address the hierarchy and triviality problems, and
for phenomenological
reasons, we are interested in new physics where $\mathcal{M} \sim \rm
TeV$.  
The resulting low
energy effective theory is the one higgs standard model supplemented
with non-renormalizable local operators, of dimension $D > 4$, which
are constructed of standard model fields invariant under the $\rm
SU(3) \times SU(2) \times U(1)$ gauge symmetry.  This approach has been applied
to precision electroweak observables\cite{Weinberg:1979sa,Wilczek:1979et, Leung:1984ni,
Buchmuller:1985jz,Grinstein:1991cd,Hagiwara:1993ck} and has recently
been the subject of further investigations\cite{Gounaris:1998ni, Plehn:2001nj,Barger:2003rs,Logan:2004hj,Han:2005pu,Manohar:2006ga,Manohar:2006gz}.
The advantage of this approach is that it is model independent: any
new physics scheme that results in a low energy spectrum coinciding
with the SM's can be described in this way. The disadvantage is that
the new physics is parametrized in terms of several arbitrary
parameters, the coefficients of higher dimension operators, and
nothing is known a priori about these coefficients.

For a particular extension to the standard model, consistency requires
that fits such as \cite{Eidelman:2004wy} be reconsidered with the new operators, severely relaxing the
constraint on the higgs mass. 
In fact, it has been shown
that the effect of higher dimension operators
\cite{Hall:1999fe,Barbieri:1999tm, Chivukula:1999az} can elminate the
mass limit on the higgs. While more exotic possibilities are
tantalizing, in this paper, we focus on the possibility that the new
physics integrated out is strongly interacting and effecting the higgs
sector above the scale $\mathcal{M}$ while the higgs itself has a
relatively low mass $m_h \lesssim 2 \,  m_t$. 

Various bounds can be placed on $\mathcal{M}$ from low energy experiments. In
particular, flavor changing neutral current bounds such as those
arising from $K^0 - \bar{K}^0$ mixing impose strong constraints,
$\mathcal{M} \geq \, 10^4 \, {\rm TeV}$.  These bounds can be
relaxed by restricting the higher dimensional operator basis
through adopting the MFV hypothesis
\cite{Dugan:1984qf,Chivukula:1987py,Hall:1990ac,D'Ambrosio:2002ex,Cirigliano:2005ck,Ali:1999we,Buras:2000dm,Bobeth:2002ch,Buras:2003jf,Branco:2006hz}.
This allows one to consider $\mathcal{M}$ to be a few $\rm TeV$ while
naturally suppressing FCNC.

However, even utilizing the MFV hypothesis to justify new physics at a
$\rm TeV$, higher dimensional corrections to the standard model could
exist that modify the relation $m_W = m_Z \, \cos \theta_W$. This
relationship is experimentally required to be respected to a fraction
of a percent. The PDG quotes $\rho_0 = 1.0002^{+ 0.0007}_{-0.0004}$ for
the global fit \cite{PDBook} of precision electro-weak observables.
This fact motivates the 
consideration of new physics being integrated out that preserves $\rho_0\approx1$
naturally, even with possible strong dynamics effecting the higgs at
the scale $\mathcal{M}$.

This can be accomplished assuming an approximate custodial $\rm SU(2)_C$ symmetry
\cite{Susskind:1978ms,Weinberg:1979bn,Sikivie:1980hm}, where the weak $ \rm
SU(2)$ gauge vector bosons  transform as a triplet and the higgs field
transforms as a triplet and a singlet.  The higgs vacum expectation
value is in the singlet representation of  $\rm SU(2)_C$ , so the
approximate symmetry is explicitly realized, and is explicitly broken
only by isospin splitting of fermion yukawa couplings and by 
hypercharge. We require that the operator
extensions to ${\mathcal{L}}_{SM}$ respect this $\rm SU(2)_{C}$
symmetry up to hypercharge and Yukawa coupling violations, as
in the standard model. Operators that break the custodial symmetry are
allowed but their coefficients are taken to be naturally suppressed. 
 
In the SM, the higgs cubic and quartic couplings are not independent
parameters, but given in terms of the higgs vacuum expectation value
$v$ and mass $m_h$. The obvious immediate effect of $D>4$ operators is
to shift all these quantities in independent ways, so that effectively
the higgs cubic and quartic couplings become independent
parameters. Of course, the shift from the SM values is somewhat
restricted, of order $(v^2/{\mathcal{M}}^2) \, C$, where $C$ is a
dimensionless coefficient, $C\sim1$. The effect on single higgs
production rates of modifications to the coupling of the higgs to weak
vector bosons or to itself was investigated in
Ref.~\cite{Barger:2003rs}. The modification of higgs decay widths 
and this general class of models was also examined in 
\cite{Giudice:2007fh}.

In this paper we address the question of whether a bound state of
higgs particles can form. In the SM a higgs bound state forms only if
the higgs is very heavy\cite{Cahn:1983vi,Rupp:1991bb}. There is a competition
between the repulsive interaction of the quartic coupling, $\lambda_1$, and
the attractive interaction of the higgs exchange (between higgs
particles) which is determined by the cubic coupling, $\lambda_1 v$. For
large enough coupling the exchange interaction is strong enough to
produce binding, but since the mass, $\sqrt{\lambda_1}v$, is also given in
terms of the coupling, the higgs mass is large.  In the effective
theory context the three parameters (mass and cubic and quartic
couplings), are independent and a bound state is possible for smaller
higgs mass. The question of detail becomes, how is the bound on the
higgs mass for a bound state to form relaxed by the coefficients of
$D=6$ operators?  Can one have a bound state of light higgses? We find
that the effect of these operators can be significant, allowing
for bound states for much lighter higgs particles. Discovery of such
bound states would give valuable information on the scale of new
physics.

There is no know solution to the bound state problem for identical
scalar particles interacting via cubic and quartic interactions.  The
higgs bound state problem has been addressed using different
approximations, the $N/D$ method is Ref.~\cite{Cahn:1983vi} and a
truncated version of the homogeneous Bethe-Salpeter equation in
Ref.~\cite{Rupp:1991bb}. Our aim here is to find a necessary condition 
on the coupling for which a non relativistic (NR) bound state may form. To this end we
introduce a new method. We propose to study the formation of the bound state in a
non-relativistic effective theory for higgs-higgs interactions.

We begin by listing the $D=6$ operators of the effective
theory. We take two approaches. In the first, linear realization, we
consider operators that can be built out of the higgs doublet and the
fields in the gauge sector of the SM. Our primary interest here is in
the higgs sector per se, so we focus on higgs self interactions. The
second approach,  uses a non-linear realization of the symmetry. Since
the higgs field is intimately connected to the symmetry breaking of the
SM gauge symmetry, it is natural to expect that below the scale of new
physics the effects of symmetry breaking are already apparent. 
Were the higgs mass as large as the scale of new physics, the SM would
be supplemented not with a higgs doublet but with a triplet of
would-be goldstone bosons that are eaten by the $W$ and $Z$ vector
bosons. The higgs, if somewhat lighter than the scale ${\cal M}$,
would appear as a singlet under the gauge symmetry.

We then proceed to construct the effective theory at low energies. If
$m_h \lesssim 2 \, m_t$, one can incorporate
the virtual effects of the top by integrating it out and constructing
a top-less effective theory. In order to investigate the minimal
coupling for which a NR  higgs-higgs bound state may form we then
construct a non-relativistic higgs effective theory, and proceed to
determine this condition.

\section{Higgs Effective Field Theory: Linear Realization}
\subsection{The $D= 6$ Custodial $\rm SU(2)$ Higgs Sector}
The Lagrangian density of the standard model containing the higgs
field\footnote{We have omitted Yukawa interactions with fermions here.} is given by
\begin{eqnarray}
\label{Lhiggs}
{\mathcal{L}}_{\phi}^4 =  
\left(D^\mu \,\phi \right)^\dagger \, \left(D_\mu \, \phi \right)  - V \left( \phi \right)
\end{eqnarray}
where $\phi$ is the higgs scalar doublet.  The covariant derivative of the
$\phi$ field is given by
\begin{eqnarray}
D_{\mu} = 1 \, \partial_\mu - i \, \frac{g_1}{2} \, B_\mu - i \, g_2 \,  \frac{\sigma^I}{2} \, W_\mu^I
\end{eqnarray}
where $\sigma^I$ are the pauli matrices, $W_\mu^I,B_\mu$, are the $\rm SU(2)$
and $\rm U(1)$ SM gauge bosons and the hypercharge of $1/2$ has been
assigned to the higgs.  The higgs potential at tree level is given by
\begin{eqnarray}
V(\phi) =  -m^2 \, \phi^\dagger \, \phi  + \frac{\lambda_1}{2} \, \left( \phi^\dagger \phi \right)^2.
\end{eqnarray}

No dimension five operator can be constructed out of higgs fields and
covariant derivatives that satisfies Lorentz symmetry and the standard
model's gauge symmetry.\footnote{To satisfy Lorentz invariance an even
number of covariant derivatives are required. To be invariant under
the $\rm SU(2) \times U(1)$ gauge group the operator must be bilinear in
$\phi^\dagger$ and $\phi$.} Utilizing the equation of motion of the higgs field
and partial integration the number of dimension six operators is
reduced.  The effective Lagrangian density of the extended standard
model is given by
\begin{eqnarray}
\mathcal{L_{\phi}} =  {\mathcal{L}}_{\phi}^4 + \frac{{\mathcal{L}}_{\phi}^6}{{\mathcal{M}}^2} +  {\mathcal{O}}(\frac{v^4}{{\mathcal{M}}^4}),
\end{eqnarray}
where the dimension six operators that preserve the symmetries of the
standard model and custodial $\rm SU(2)_C$ in the Higgs sector are given by
\begin{eqnarray}
{\mathcal{L}}_{\phi}^6 &=&  C_{\phi}^1 \, \partial^\mu \, ( \phi^\dagger \, \phi) \partial_\mu \, ( \phi^\dagger \, \phi)
+ C_{\phi}^2 \, \left( \phi^\dagger \, \phi \right) \,  \left(D_\mu \, \phi \right)^\dagger \, \left(D^\mu \, \phi \right) -  \frac{\lambda_2}{ 3 \, ! } \, \left( \phi^\dagger \, \phi \right)^3. 
 \end{eqnarray}



Note that the operators considered here preserve custodial symmetry and
can result from tree level topologies in the underlying theory.\cite{Arzt:1994gp} As such, these operators
need not be suppressed by loop factors of $1/16 \pi^2$ or proportional to a
small custodial symmetry breaking parameter. For these reasons these operators are
expected to have the dominant effects on the higgs self couplings and we take their coefficients
to be $\mathcal{O} (1)$.  There is only one operator in the Higgs sector that violates custodial symmetry and could come from an underlying tree topology,  $(\phi^\dagger \, D^\mu \, \phi)^2$. The
underlying topology in this case determines that the symmetry breaking parameter
is given by $g_1^2$. The coefficient of this operator has been determined \cite{Barbieri:2004qk} to be
$ C < 4 \times 10^{-3}$ where we have used $\Lambda = 1 \, {\rm TeV}$. We neglect this operator.

We expand the higgs field about its vacuum expectation value with $\langle
h(x) \rangle = 0$ and treat $v^2/ \mathcal{M}^2$ as a small perturbation.
We expand the field as usual around a vacuum expectation value $v$ so
that
\begin{eqnarray}
\phi (x) = \frac{{\rm  U(x)}}{\sqrt{2}} \,
\left(
 \begin{array}{c} 
 0 \\
v + h(x) 
\end{array}
\right).
\end{eqnarray}
Here $ {\rm U(x)} = e^{i \, \xi^a(x) \, \sigma_a/v}$ and the would-be
goldstone boson fields of the broken symmetry are $\xi^a$.
In unitary gauge, the  gauge transformation can be used
to remove the goldstone boson fields. We then redefine the  higgs
field ($h$) so that the kinetic term is normalized to $1/2$, using the
field redefinition
\begin{eqnarray}
h \to \frac{h'}{(1+ 2 \, C_h^K)^{1/2}},
\end{eqnarray} 
where $C^K_h= (v^2/ \mathcal{M}^2)(C^1_\phi+\tfrac14C^2_\phi )$. The effective
Lagrangian density is given, in terms of the rescaled field, by
\begin{eqnarray}
\label{Leffhprime}
{\mathcal{L}}_\phi^4 + \frac{{\mathcal{L}}_\phi^6}{\mathcal{M}^2} &=&   \frac{1}{2}\, \partial^{\mu} \, h' \, \partial_\mu \, h' - V_{eff}(h') + C_{h'}^{i,j} \, O_{h'}^{i,j}
+ C_{W \, W} \, O_{W \, W} + C_{Z \, Z} \, O_{Z \, Z} \nn \\
&\,&+\, C_{h' \, W \, W}^{i,j} \, O_{h' \,W \, W}^{i,j}
+ C_{h' \, Z \, Z}^{i,j} \, O_{h' \, Z \, Z}^{i,j},
\end{eqnarray}
summed over $i,j$ such that $i+ j = 2$ , where
\begin{align} 
  O_{h'}^{i,j} &= \frac{(h')^i \, v^j}{\mathcal{M}^2} \, \partial^{\mu} \, h' \, \partial_\mu \, h', \nn\\
  O_{W \, W} &= W^{+}_{\mu} \, W^{-}_{\mu} \,,
&
  O_{Z \, Z} &= Z^{0}_{\mu} \, Z^{0}_{\mu},  \nn\\
  O_{h' \, W \, W}^{i,j} &=  \frac{(h')^i \, v^j}{\mathcal{M}^2} \, W^{+}_{\mu} \, W^{-}_{\mu},
 &
  O_{h' \, Z \, Z}^{i,j} &= \frac{(h')^i \, v^j}{\mathcal{M}^2} \,Z^{0}_{\mu} \, Z^{0}_{\mu}.
\end{align}
The coefficients are given by
\begin{align} 
 C_{h'}^{0 \, , 2} &= 0,
& C_{h'}^{1 \, , 1} &= \frac{1}{2} \, \left( 4 \, C_{\phi}^1 + C_{\phi}^2 \right),  \nn\\
  C_{h'}^{2 \, , 0} &= \left( C_{\phi}^1 + \tfrac14C_{\phi}^2 \right) \,,
&
 C_{W \, W}&= m_W^2 \, \left(1+ C_{\phi}^2 \, \frac{v^2}{2 \, \mathcal{M}^2} \right), \nn\\
 C_{Z \, Z} &= \frac{m_Z^2}{2} \, \left(1+ C_{\phi}^2 \, \frac{v^2}{2 \, \mathcal{M}^2} \right), 
 &
    C_{h' \, W \, W}^{1\, , 1}&=  m_W^2 \, \left[\frac{3}{2} \, C_{\phi}^2 - 2 \, C_{\phi}^1 + \frac{2 \, \mathcal{M}^2}{v^2} \right], \nn\\
    C_{h' \, W \, W}^{2\,, 0} &= m_W^2 \, \left[\frac{5}{2} \, C_{\phi}^2  - 2 \, C_{\phi}^1 + \frac{ \mathcal{M}^2}{v^2} \right],  
  &
      C_{h' \, W \, W}^{3\,, -1}&= 2 \, m_W^2 \, C_{\phi}^2,  \nn\\
      C_{h' \, W \, W}^{4\,, -2} &= \frac{1}{2} \, m_W^2 \, C_{\phi}^2,
 &
  C_{h' \, Z \, Z}^{1\, , 1}&= \frac{m_Z^2}{2} \, \left[\frac{3}{2} \, C_{\phi}^2 -  2 \, C_{\phi}^1 + \frac{2 \mathcal{M}^2}{v^2} \right],  \nn\\
  C_{h' \, Z \, Z}^{2\,, 0} &= \frac{m_Z^2}{2} \, \left[\frac{5}{2} \, C_{\phi}^2  - 2 \, C_{\phi}^1 + \frac{ \mathcal{M}^2}{v^2} \right], 
  &
C_{h' \, Z \, Z}^{3\,, -1}&= m_Z^2 \, C_{\phi}^2,    \nn\\
C_{h' \, Z \, Z}^{4\,, -2} &= \frac{m_Z^2}{4} \, C_{\phi}^2.
\end{align}

 The effective potential is
\begin{eqnarray}
\label{Veffhprime}
V_{eff}(h')  = \frac{1}{2} \, m_h^2 h'^2 + \frac{v \, \lambda_3^{eff}}{3 \, !} \, h'^3 + \frac{\lambda_4^{eff}}{4 \, !} \, h'^4 + \frac{30 \, \lambda_2}{5 \, !  \, \mathcal{M}^2} \, v \,  h'^5 + \frac{30 \, \lambda_2}{6 \, ! \, \mathcal{M}^2} \,  h'^6, 
\end{eqnarray} 
which is written in terms of the rescaled mass term and the effective
couplings, which are given by
\begin{eqnarray}
\label{effmass1}
\frac{m_h^2}{v^2} &=& \lambda_1\, \left(1 - 2 \, C_{h}^K \right)+
\frac{\lambda_2}{2} \, \frac{v^2}{\mathcal{M}^2}  +
     {\mathcal{O}}(\frac{v^4}{\mathcal{M}^4}), \\
\label{lambda3eff1}
\lambda_3^{eff}&=& 3 \, \lambda_1\, \left(1 - 3 \, C_{h}^K \right)  + \frac{5}{2} \, \lambda_2 \, \frac{v^2}{\mathcal{M}^2} + {\mathcal{O}}(\frac{v^4}{\mathcal{M}^4}), \\
\label{lambda4eff1}
\lambda_4^{eff}&=& 3 \, \lambda_1 \, \left(1 - 4 \, C_{h}^K \right)  +  \frac{15}{2} \, \lambda_2 \, \frac{v^2}{\mathcal{M}^2}
+ {\mathcal{O}}(\frac{v^4}{\mathcal{M}^4}). 
\end{eqnarray} 
We will
suppress the prime superscript on the higgs field for the remainder of
the paper.

\subsection{$D=6$ SM Field Strength Operators}
The operators that can be constructed out of the higgs scalar doublet and
the field strengths (or duals) of the standard model are as follows.  We
restrict our attention to those operators listed in
\cite{Buchmuller:1985jz,Manohar:2006gz} that preserve the  $\rm
SU(2)_C$ custodial symmetry: 
\begin{eqnarray}
\label{HHGG}
\frac{{\mathcal{L}}_{\phi, V}^6}{{\mathcal{M}}^2} &=& -  \frac{c_G \, g_3^2}{2 \, {\mathcal{M}}^2} \, \left(\phi^\dagger \, \phi \right) \, G^A_{\mu \, \nu} \, G^{A \, \mu \, \nu} 
- \frac{c_W \, g_2^2}{2 \, {\mathcal{M}}^2} \, \left(\phi^\dagger \, \phi \right) \, W^I_{\mu \, \nu} \, W^{I \, \mu \, \nu}
- \frac{c_B \, g_1^2}{2 \, {\mathcal{M}}^2} \, \left(\phi^\dagger \, \phi \right) \, B_{\mu \, \nu} \, B^{\mu \, \nu}, \nn  
\\
&\,& - \frac{\tilde{c}_{G} \, g_3^2}{2 \, {\mathcal{M}}^2} \, \left(\phi^\dagger \, \phi \right) \, \tilde{G}^A_{\mu \, \nu} \, G^{A \, \mu \, \nu}
 - \frac{\tilde{c}_{W} \, g_2^2}{2 \, {\mathcal{M}}^2} \, \left(\phi^\dagger \, \phi \right) \, \tilde{W}^I_{\mu \, \nu} \, W^{I \, \mu \, \nu} \nn \\
 &\,& - \frac{\tilde{c}_{B} \, g_1^2}{2 \, {\mathcal{M}}^2} \, \left(\phi^\dagger \, \phi \right) \, \tilde{B}^A_{\mu \, \nu} \, B^{A}_{\mu \, \nu}.
\end{eqnarray} 
Here $G^A_{\mu \, \nu}$, $W^I_{\mu \, \nu} $ and $B_{\mu \, \nu}$ stand for the
field strength tensors of the $SU(3)\times SU(2)\times U(1)$ gauge bosons, and a
tilde denotes the dual field strengths, $\tilde{F}_{\mu \, \nu} = \epsilon_{\mu \,
\nu \, \lambda \, \sigma} \, F^{\lambda \, \sigma} \, /2$.
Note that the operator that is proportional to the $S$ parameter given by
\begin{eqnarray}
- \frac{c_{W \, B} \, g_1 \, g_2}{{\mathcal{M}}^2} \, \left(\phi^\dagger \, \sigma^I \, \phi \right) \, B^{\mu \, \nu} \, W_{I \, \mu \, \nu} 
\end{eqnarray} 
violates custodial symmetry and is naturally suppressed in our approach \footnote{See Appendix A}.

\subsection{$D=6$ Fermion Sector}

Operators of dimension 5 and higher that couple the higgs to fermions,
or purely fermionic operators, can give rise to unacceptably large
flavor changing neutral currents (FCNC). If the coefficient of such
operators are generically of order 1 the scale of new physics must be
taken to be $\mathcal{M}\gtrsim 10^4$~TeV in order to suppress 
FCNC effects. We adopt the Minimal Flavor Violation hypothesis (MFV) to
naturally suppress the dangerous operators while maintaining a low
scale of new physics, $\mathcal{M}\gtrsim 1$~TeV.  In the absence of quark
and lepton masses the SM has a large flavor symmetry group,
$G_F=SU(3)^5$. The  MFV asserts that there is a unique source of
breaking of this symmetry. All operators that break the symmetry must
transform precisely the same way under $G_F$. As a result  FCNC
operators are suppressed by the familiar factors of the
Kobayashi-Maskawa (CKM) matrix in the quark sector and by the  
Pontecorvo-Maki-Nakagawa-Sakata (PMNS) matrix and small neutrino
masses in the lepton sector. 

Since the effects of fermionic operators are not needed for the rest
of this investigation, we do not list the operators. The interested
reader can find a complete description of the operators and their
effects in \cite{Buchmuller:1985jz}.

\section{Higgs Effective Field Theory: Non Linear Realization}
The construction in the previous section assumes that the field
content of the effective theory includes a higgs doublet. This is not
necessary. If the electroweak symmetry is spontaneously broken by a
strong interaction the spectrum below the scale of this new physics
does not have to be described by a higgs doublet field, beyond the SM
fields. Only fields describing the would-be goldstone bosons need be
introduced. Such higgs-less theories have been discussed in the
literature\cite{Appelquist:1993ka}. However, if the higgs particle is
somewhat lighter than the scale of new physics it has to be
incorporated in the low energy description and symmetry alone does not
dictate that it appears as a member of an iso-doublet. It is
sufficient to have the goldstone bosons realize the broken symmetry
non-linearly, and the higgs field is then a singlet under the
symmetry.

The situation is entirely analogous to the case of $\pi$'s and the
$\sigma$ in QCD. A phenomenological Lagrangian density describing $\pi$ and
$\sigma$ interactions does not have to be a linear realization of the
chiral $SU(2)\times SU(2)$ symmetry. Instead, the $\pi$-fields have a better
description through a non-linear
chiral Lagrangian. Then the $\sigma$ can be included through interactions
that satisfy the non-linearly realized symmetry and the usual rules
for naive dimensional analysis \cite{Manohar:1983md}. 

In the non-linear realization, the Lagrangian density in Eq.~\eqref{Lhiggs} is
replaced by
\begin{equation}
\mathcal{L}_{\rm NL} =\tfrac14 v^2\text{Tr}D_\mu{\rm U}^\dagger D^\mu{\rm U} 
+\tfrac12 \partial_\mu h\partial^\mu h -V(h),
\end{equation}
where the would-be goldstone bosons $\xi^a$ appear through the matrix
$ {\rm U(x)} = e^{i \, \xi^a(x) \, \sigma_a/v}$ that transforms under $ \rm  SU(2)_L\times SU(2)_R$
linearly, ${\rm U} \to L{\rm U} R^\dagger $, and $h$ is a singlet field, describing the
higgs particle. Custodial symmetry $\rm SU(2)_C$ is the diagonal subgroup
of $\rm SU(2)_L\times SU(2)_R$ and the higgs field is invariant under it.
\footnote{The custodial symmetry is discussed in more detail in Appendix A.}

This Lagrangian is supplemented by higher order terms suppressed by
powers of $\mathcal{M}$. In the case of the higgs potential, this can
be included simply  as
\begin{equation}
V(h)=\mathcal{M}^4f(h/ \mathcal{M}),
\end{equation}
where $f(x)$ is an arbitrary function with a minimum at zero. The mass and
couplings of the higgs are given in terms of this dimensionless function 
by
\begin{align}
m_h^2&=\mathcal{M}^2f^{(\prime\prime)}(0),\\
\label{eq:BIG}
v\lambda_3^{eff}&=\mathcal{M}f^{(\prime \prime \prime )}(0),\\
\lambda_4^{eff}&=f^{(iv )}(0).
\end{align}
It is not a surprise that in the non-linear realization of the
symmetry the couplings and mass are completely independent, and that
they are all naturally of order 1 times the appropriate power of the
dimensionfull scale, $\mathcal{M}$. The natural scale for the higgs
mass is $\mathcal{M}$, and we are considering here the class of
theories for which $f^{(\prime\prime)}(0)$ happens to be small, while higher
derivatives may remain of order 1. We stress that the natural scale
for the cubic coupling is $\mathcal{M}$. {\it Unless the mechanism (or
numerical accident) that keeps the higgs mass small compared to
$\mathcal{M}$ also acts to suppress the cubic coupling, one must
naturally expect $\lambda^{eff}_3\sim\mathcal{M}/v \gg1$.}

We will also need  the corrections to the derivative interactions. We write, generally,
\begin{equation}
\mathcal{L}= \frac{1}{2}\left[1+c_1^{eff}\frac{h}{v}+c_2^{eff}\frac{h^2}{v^2}\right]
 \, \partial^\mu \, h \,  \partial_\mu \, h - \frac{1}{2} \, m_h^2 \, h^2 - \frac{v \, \lambda^{eff}_3}{3 \, ! } \, h^3  - \frac{\lambda^{eff}_4}{4 \, ! } \, h^4 +  \cdots
\end{equation}
In the linear realization the derivative interaction couplings are
related, $\tfrac14c_1^{eff}=\tfrac12c_2^{eff}=C_h^K=(v^2/
\mathcal{M}^2)(C_\phi^1+C_\phi^2/4)$, but in the non-linear realization they
are independent. And, as in the case with $\lambda_3^{eff}$ naive
dimensional scaling gives an enhancement of $c_1^{eff}$ that could
arise from the non-perturbative dynamics of the symmetry breaking
sector. Naively, $c_1^{eff}\sim(v/ \mathcal{M})$, which is enhanced
over the linear realization value by a power of $(\mathcal{M}/v)$.

As we mentioned earlier, non-linear realizations have been extensively
studied for higgs-less theories, but have been neglected in  studies  including
a light higgs. There are two important consequences of the
non-linear realization outside the pure higgs sector that we point out
here. It has been noted that significant corrections to the coupling
of a higgs to gluons are possible from $D>4$ operators. The
modifications can be large because there is no SM contribution at tree
level. In the linear realization there is a $D=6$ operator that
contributes at tree level, and therefore competes with the SM one
loop, top mediated amplitude:
\begin{equation}
\label{GGeff}
\frac1{\mathcal{M}^2}G^a_{\mu\nu}G^a_{\mu\nu},
\end{equation}
Note that the linear realization implies a relation between the one
and two higgs couplings to two gluons. However, in the non-linear
realization the two couplings are completely independent,
\begin{equation}
\label{hvshh}
\left(c_1\frac{h}{\mathcal{M}}+c_2\frac{h^2}{\mathcal{M}^2}\right)G^a_{\mu\nu}G^a_{\mu\nu}.
\end{equation}
In Ref.~\cite{Pierce:2006dh} it was noted that  a heavy quark with 
Yukawa coupling $\lambda \to \infty$ produces a
coupling of two gluons to one or more higgs particles that cannot be
described  by the effective theory operator in \eqref{GGeff}. 
Instead a nonpolinomial interaction was introduced to describe this effect, 
\[
\frac{\alpha_s}{8\pi}\ln(\frac{H^\dagger H}{v^2})G^a_{\mu\nu}G^a_{\mu\nu}.
\]
There is
no problem accommodating  such interactions in the non-linear
realization, 
by 
\begin{equation}
\label{eq:pierce}
\frac{\alpha_s}{4\pi}\ln(1+h/v)G^a_{\mu\nu}G^a_{\mu\nu}=\frac{\alpha_s}{4\pi}[h/v+(h/v)^2+\cdots]G^a_{\mu\nu}G^a_{\mu\nu}.
\end{equation}

In much of what follows we implicitly assume the linear
realization. However, results in terms of the arbitrary parameters
$m_h$, $\lambda_3^{eff}$ and $\lambda_4^{eff}$ can be interpreted readily as
arising from the non-linear realization.

\section{A Low Energy Effective Theory for the higgs}
In this section we will construct an effective theory for the light
higgs, integrating out momentum modes heavier than the higgs. This is
useful in discussing physical effects with a typical energy of order
of the higgs mass. In particular, we integrate out the top
quark.
As the coupling of the top quark to the higgs is fairly large, we would
like to estimate the effects of the top quark on the possibility of
forming a higgs bound state.  If the top quark mass is much heavier than the
higgs  it is appropriate and convenient to describe the higgs self
interactions in a top-less theory. When the top quark has been
integrated out, its effects are accounted for through modifications of
coupling constants and mass of the higgs. 

While this is clearly appropriate when the top quark is much
larger than the higgs mass, we use this approximation even
when the higgs is slightly heavier than the top. For $m_h < 2 \, m_t$ the
approximation is known, {\it ipso facto}, to  work better than 
one would expect. This is due, in part, to the fact that there
is no non-analytic dependence on the mass since the higgs is the
pseudo-goldstone boson of spontaneously broken scale
invariance\cite{Grinstein:1988yu,Dawson:1989yh,Chivukula:1989ze}.
It is also known that soft gluon effects are large and correctly reproduced by 
the effective theory \cite{Kramer:1996iq}.

For single and double higgs production, comparisons between the 
the full theory calculation and the effective top-less theory find that 
the latter is a good approximation for the total rate
for $m_h \lesssim 2 \, m_t$. 
For example, with the appropriate K factor, the resulting
topless effective field theory calculated to two loops is known to
accurately describe the full NLO result for $g \, g \to h$ to better
than $5\%$ accuracy in the full range $0<m_h < 2 \, m_t $
\cite{Kramer:1996iq}. 

As another concrete example, consider the higgs mass dependence in 
the higgs IPI self-energy. The first graph of
Fig.~\ref{fig:topout} is the contribution of the
top quark to the IPI self-energy which we label $- i \, \Pi(p^2)$.
 At 1-loop we find 
\begin{eqnarray}\label{senergy}
 \Pi(p^2) =  \frac{N_c}{4 \pi^2} \frac{m_t^4}{v^2} \, \int_0^1 \, {d \, x} \, \left(1-x(1-x)\frac{p^2}{m_t^2}\right) \left[1 + 3 \, \log \left(\frac{\mu^2}{ m_t^2-x(1-x)  p^2 }\right) \right], 
\end{eqnarray}
where we have used the $\rm \overline{MS}$ subtraction scheme.  The
quantity $1-\Pi(0)/\Pi(p^2)$ (at $\mu=m_t$) never exceeds 30\%  when
$\sqrt{p^2}$ ranges from zero to $2 \, m_t$. 

For these reasons we consider it appropriate to integrate out the top quark for $m_h \lesssim 2 \, m_t$
in this initial study. When $m_h >>  m_t$,  these corrections should be taken only as
indications of the size of virtual top effects. While the approximation of neglecting higher 
order terms in the $p^2/m_t^2$ expansion is known to work better than expected for the applications
we will consider, there is  no guarantee that it will work well for processes not considered here.
\cite{Cohen:1983fj,D'Hoker:1984ph,Feruglio:1992fp,Lin:1991jr,Lin:1993gx} 

\subsection{Running to $m_t$} 

The coefficients of the $D=6$ operators at the scale $\mathcal{M}$ are
unknown.  We are assuming
that the new physics couples to the higgs field and is strongly
interacting at the scale $\mathcal{M}$. In
this context, it is natural to take 
\begin{eqnarray}
C_{\phi}^i \left(\mathcal{M}\right), \, \lambda_2 \left(\mathcal{M}\right) \sim 1.
\end{eqnarray}
Similarly it is natural to assume that the coefficients of the $D>4$
operators that couple the higgs to other fields, like those in
Eq.~\eqref{HHGG} or those that couple the higgs to quarks while
satisfying the MFV hypothesis, are all order unity.

The anomalous dimensions of the extended operator basis can be
determined systematically.  This is beyond the scope of this
paper. But the effect of the running is easy to understand. With
minimal subtraction  the
calculation of the running of  coefficients of higher dimension
operators can be done in the symmetric, massless phase. There is
operator mixing among the $D=6$ operators with common quantum
numbers. The anomalous dimension matrix is a function of the
relevant couplings ($\lambda_1$, $g_1$, $g_2$ and  the top quark Yukawa, $\lambda_t$).
The running is always proportional to these
coefficients so the effect is roughly of the form
\begin{eqnarray}
C_{\phi}^i \left( m_t \right)\sim C_{\phi}^i \left( \mathcal{M} \right) \,
 \left(1 + \frac{c_1\alpha}{16\pi^2}  \,  \log \left(\frac{m_t}{\mathcal{M}}
 \right)  \right), 
\end{eqnarray}
where mixing is implicit, and $c_1\alpha$ stands for a linear
combination of $\lambda_1$ and the squares of $g_1$, $g_2$ and  $\lambda_t$. 

Since $ \log \left(m_t / \mathcal{M} \right)\sim 1$ and
the coefficients $c_1 \sim1$ the running produces a small, calculable  shift in the
unknown coefficients. Hence, we continue to take the unknown Wilson
coefficients at the scale $m_t$ to be $\sim 1$ .  

At $m_t$ the top quark
is integrated out and this produces a different effect, a shift in the
$C^i_\phi(m_t)$ by a $C^i_\phi$-independent amount. This can be numerically
 significant, and we estimate this next.  Note that once the top is
 integrated out we continue to run down to the mass of the higgs
scalar $m_h$. The effect of the running of these coefficients
from $m_t$ to $m_h$ is again small, so we take
\begin{eqnarray}
C_{\phi}^i \left(m_h \right), \, \lambda_2 \left( m_h \right) \sim 1.
\end{eqnarray}

\subsection{Integrating out the top quark}

\begin{figure}[hbt]
\centerline{\scalebox{0.8}{\includegraphics{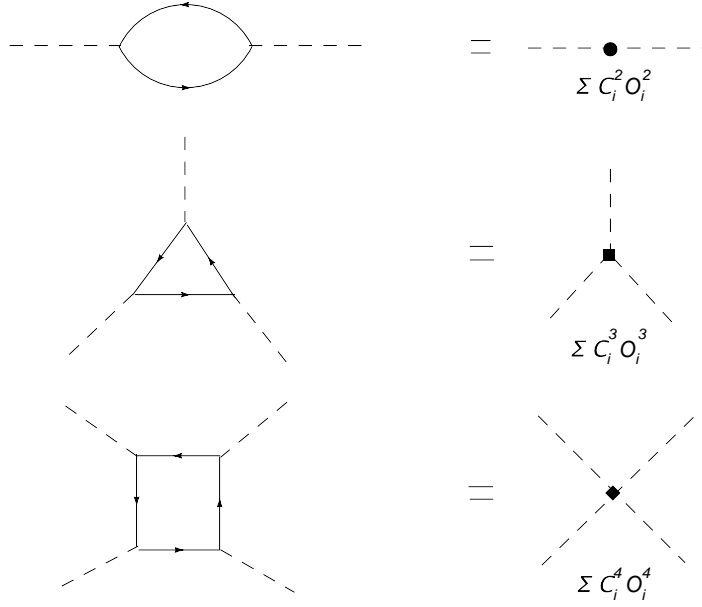}}}
\caption{Integrating out the top quark.}
\label{fig:topout}\end{figure}

Integrating out the top leads to further corrections to the higgs
sector of the standard model. The top mass is a result of symmetry
breaking, so the resulting effective theory is better presented in
unitary gauge, as in \eqref{Leffhprime} and \eqref{Veffhprime}. In
unitary gauge, the top mass term and coupling to the higgs is given by
\begin{eqnarray}
\mathcal{L}_Y = - m_t  \, \bar{q}_t \, q_t \, \left(1+ \frac{h}{v} \right).
\end{eqnarray}
We begin by considering effects on the higgs
self-couplings. Figure~\ref{fig:topout} show the Feynman graphs
that contribute to modifications of the higgs self-couplings. The solid
line denotes a top quark, the dashed external lines denote the higgs.

We perform the calculation to lowest order in $p^2/m_t^2$. 
Some details of the computation are given in the appendix.
The effect of these corrections is to further modify the effective potential of the higgs scalar field $h$. 
The effective couplings and mass term of
Eqs.~\eqref{effmass1}--\eqref{lambda4eff1} are shifted by  these
corrections, and are now are given by
\begin{eqnarray}
\frac{m_h^2}{v^2} &=&\lambda_1 \left(1 - 2 \, C_{h}^K \right)  + \frac{N_c}{4 \, \pi^2} \, \left( \frac{m_t^4}{v^4} \right) + \frac{\lambda_2}{2} \, \frac{v^2}{\mathcal{M}^2}  + {\mathcal{O}}(\frac{v^4}{\mathcal{M}^4},\frac{m_t^2 \, m_h^2}{v^4}), \\
\lambda_3^{eff}&=&3 \, \lambda_1 \, \left(1 - 3 \, C_{h}^K \right) - \frac{N_c}{\pi^2} \, \left( \frac{m_t^4}{v^4} \right)  + \frac{5}{2} \, \lambda_2 \, \frac{v^2}{\mathcal{M}^2} + {\mathcal{O}}(\frac{v^4}{\mathcal{M}^4},\frac{m_t^2 \, m_h^2}{v^4}), \\
\lambda_4^{eff}&=& 3 \, \lambda_1  \, \left(1 - 4 \, C_{h}^K \right)  - \, \frac{4N_c}{\pi^2} \, \left( \frac{m_t^4}{v^4} \right) +  \frac{15}{2} \, \lambda_2 \, \frac{v^2}{\mathcal{M}^2}
 + {\mathcal{O}}(\frac{v^4}{\mathcal{M}^4},\frac{m_t^2 \, m_h^2}{v^4}). 
\end{eqnarray} 

As emphasized above, these corrections are not multiplicative, that
is, they are present even for $\lambda_2=C_h^K=0$. Whether they are
important depends on the scale and strength of the new physics. The condition
\begin{eqnarray}
\lambda_2 \, \frac{v^2}{{\mathcal{M}^2}}  \sim \left(\frac{m_t^4}{v^4}\right) \frac{1}{\pi^2}.
\end{eqnarray}
is satisfied for $\lambda_2\approx 1$ when $\mathcal{M}\approx2\pi v=1.6$~TeV.  So the
corrections are numerically comparable to these new physics
terms. Similarly, for $\lambda_1\sim1$ the condition
\begin{eqnarray}
C_h^K \lambda_1=\frac{v^2}{\mathcal{M}^2}(C_\phi^1+\tfrac14C_\phi^2)\lambda_1   \sim \left(\frac{m_t^4}{v^4}\right) \frac{1}{\pi^2}.
\end{eqnarray}
still requires $\mathcal{M}\approx1.6$~TeV for $C_\phi^1+\tfrac14C_\phi^2\sim1$.

\subsection{Corrections to Field Strength operators}\label{twohiggsprod}

\begin{figure}[hbtp]
\centerline{\scalebox{1}{\includegraphics{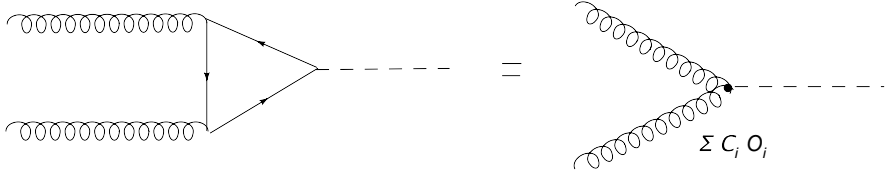}}}
\caption{The gluon fusion $g \, g \to h$ production process. The production process through the effective local operators in shown in the second column. The effective local operators come from integrating out the top quark and new physics at $\mathcal{M}$.}
\label{gluonfusiontri}
\end{figure}

Integrating out the top quark also results in effective operators of
the higgs field and the SM field strengths. The dominant SM production
mechanisms for the higgs at LHC is the gluon fusion process $g \, g \to
h$. We restrict our attention to such operators that effect the
production processes of the higgs through gluon fusion.  Figure~\ref{gluonfusiontri}
shows the 1-loop Feynman diagram for the top contribution to $g \, g \to
h$. For a higgs
with $m_h < 2 \, m_t$, the expected production cross section of the $g \, g
\to h$ process has been determined up to NNLO \cite{Harlander:2002wh,
Anastasiou:2002yz, Ravindran:2003um}. For SM gluon fusion, the single
higgs production mechanism is given by the $m_t \to \infty$ effective
Lagrangian density comprised of a dimension five operator
\begin{eqnarray}
\mathcal{L}_{m_t} = C_{G \, G \, h}^1 \,  (\alpha_s) \, \frac{h}{v} \, G^a_{\mu \, \nu} \, G^{\mu \, \nu}_a, 
\end{eqnarray}
where the  coefficient is given in the $\rm \overline{MS}$
scheme, in terms of $\alpha_s$ for five active flavors, by
\cite{Shifman:1979eb, Vainshtein:1980ea, Voloshin:1985tc,
Dawson:1990zj,Chetyrkin:1997sg,Chetyrkin:1997un}
\begin{eqnarray}
C_{G \, G \, h}^1 \, (\alpha_s) =  \frac{\alpha_s}{12 \, \pi } +  \frac{11 \, \alpha_s^2}{48 \, \pi^2 } + \mathcal{O} (\alpha_s^3).
\end{eqnarray}

The interactions in the effective Lagrangian of Eq.~\eqref{HHGG} also
contribute to single higgs production through gluon fusion. Combining
results, at the scale $m_h$, the effective Lagrangian density for
single higgs production is given by
\begin{eqnarray}
\mathcal{L}^{eff} =  C_{G \, G \, h}^{eff} \,  \frac{h}{v} \, G^A_{\mu \, \nu} \, G^{A \, \mu \, \nu} +  \tilde{C}_{G \, G \, h}^{eff} \,   \frac{h}{v} \, \tilde{G}^A_{\mu \, \nu} \, G^{A \, \mu \, \nu},
\end{eqnarray}
where
\begin{eqnarray}
C_{G \, G \, h}^{eff} &=& C_{G \, G \, h}^{1} -   2 \,  \pi \, \alpha_s \,c_G \,\frac{ v^2 }{{\mathcal{M}}^2},  \\
\tilde{C}_{G \, G \, h}^{eff} &=&  -2 \,  \pi \, \alpha_s \, \tilde{c}_{G} \,   \frac{v^2 }{{\mathcal{M}}^2}.
\end{eqnarray}

Assuming that the new physics degrees of freedom carry the $\rm SU(3)$
gauge charge, the Wilson Coefficients $c_G, \tilde{c}_G$ will be
approximately the same size as the coefficients $C_{\phi}^i, \lambda_2$ we are
interested in.  If the new physics degrees of freedom are charged
under $SU(2) \times U(1)$ but not $\rm SU(3)$, below the scales
$\mathcal{M},m_t$ effective local operators of this form will still be
induced. However, the corresponding Wilson Coefficients will be
suppressed by factors of $16 \, \pi^2$.

The effect of these interactions on higgs production rates was
examined in \cite{Manohar:2006ga}.  Note that in the standard model,
contributions to the operator $ \tilde{G}^A_{\mu \, \nu} \, G^{A \, \mu \,
\nu}$ are highly suppressed\cite{Ellis:1978hq}
and therefore
neglected.

\begin{figure}[hbtp]
\centerline{\scalebox{1}{\includegraphics{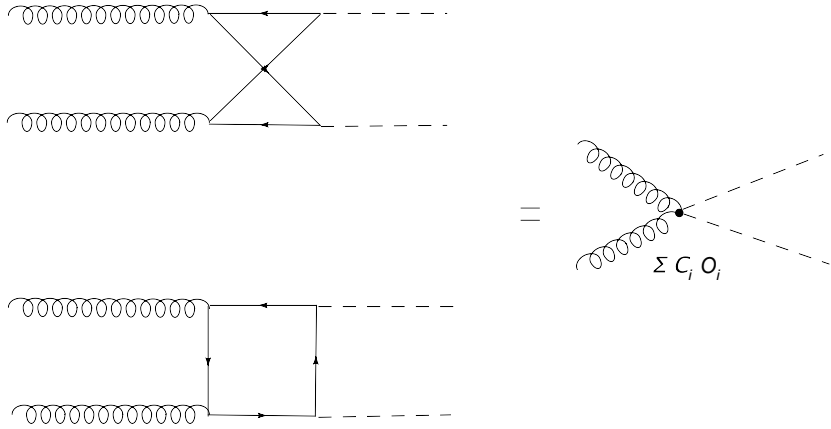}}}
\caption{The gluon fusion $g \, g \to h \, h$ production process and the
effective local operators.}
\label{gluonfusionbox}
\end{figure}

The production process of two higgs in the standard model is shown in 
Fig.~\ref{gluonfusionbox}. In analogy with the single higgs production
case we characterize the process in the effective theory by
\begin{eqnarray}
\mathcal{L}^{eff} =  C_{G \, G \, h \, h}^{eff} \,  \frac{h^2}{v^2} \, G^A_{\mu \, \nu} \, G^{A \, \mu \, \nu} +  \tilde{C}_{G \, G \, h \, h}^{eff} \,   \frac{h^2}{v^2} \, \tilde{G}^A_{\mu \, \nu} \, G^{A \, \mu \, \nu},
\end{eqnarray}
where the coefficients are given by
\begin{eqnarray}
C_{G \, G \, h \, h}^{eff} &=&  C_{G \, G \, h \, h}^{1} -  \frac{c_G \,   \pi \, \alpha_s \, v^2 }{{\mathcal{M}}^2},  \\
\tilde{C}_{G \, G \, h \, h}^{eff} &=&  - \frac{\tilde{c}_{G} \,  \pi \, \alpha_s \, v^2 }{{\mathcal{M}}^2}.
\end{eqnarray}
Here the top quark
contribution is\cite{Dawson:1990zj}
\begin{eqnarray}
C_{G \, G \, h \, h}^1(\mu^2) =  - \frac{\alpha_s(\mu^2)}{12 \, \pi} -  \frac{11 \, \alpha_s^2}{48 \, \pi^2 } + \mathcal{O} (\alpha_s^3).
\end{eqnarray}
The Wilson coefficients for two higgs production in the effective theory is not
suppressed relative to the corresponding Wilson coefficient for 
single higgs production. Note that unlike the case of single higgs production
the expansion in $p^2/m_t^2$ does not, in general,  have kinematics such that 
$p^2/m_t^2 \sim m_h^2/m_t^2$. In two higgs production, higher order
terms in $p^2/m_t^2$ have $p^2 = s,t,u$ and in general $(s,t,u)/m_t^2$ is not small.   
We calculate the next order in the expansion of $p^2/m_t^2$ in Appendix \ref{opeappend}. 
These terms are neglected, and our application of the expansion is valid for finite values of $m_t$ 
due to our interest in establishing a necessary condition for a NR bound state to form. 
The kinematics for the production of a NR bound state at threshold dictate $(s,t,u)/m_t^2 \sim m_h^2/m_t^2$.

\section{Phenomenology of  Higgs Effective Theory}

\subsection{The Magnitude of Self Couplings}

The effect of the $D=6$ operators in the effective potential cause
corrections to the three and four point contact interactions and
$m_h$.  To illustrate that the induced effects on the higgs sector are under
control, consider extending the effective potential with a single dimension
eight term.
We find the following while neglecting the effects of integrating out the top quark
\begin{eqnarray}
\frac{m_h^2}{v^2} &=&  \lambda_1\, \left(1 - 2 \, C_{h}^K \right)^2 +  3 \times 10^{-2}  \, \lambda_2 \, \left(1 - 2 \, C_{h}^K \right)  + 4.5 \times 10^{-4} \, \lambda_3, \\
\lambda_3^{eff}&=& 3 \,  \lambda_1\, \left(1 - 3 \, C_{h}^K + 7.5 \, (C_{h}^K)^2 \right) + 1.5 \times 10^{-1}  \, \lambda_2 \,  \left(1 - 3 \, C_{h}^K \right)+ 3.1 \times 10^{-3} \, \lambda_3, \\
\lambda_4^{eff}&=& 3 \,  \lambda_1 \, \left(1 - 4 \, C_{h}^K + 12 \, (C_{h}^K)^2 \right) +4.5 \times 10^{-1}  \, \lambda_2 \, \left(1 - 4 \, C_{h}^K \right) + 1.6 \times 10^{-2}  \, \lambda_3. 
\end{eqnarray} 
From which one sees we are examining the potential of the theory in a controlled expansion, even for 
$\mathcal{M} \sim 1 \, \rm TeV$.

Eliminating the self-coupling $\lambda_1$ in favor of the higgs mass, we
can write for the effective cubic and quartic higgs-self couplings,
\begin{eqnarray}
\lambda_3^{eff} = 3 \, \left(1- C_{h}^{K} \right)\frac{m_h^2}{v^2}
+\lambda_2\frac{v^2}{\mathcal{M}^2} - \frac{7N_c}{4  \, \pi^2} \, \frac{m_t^4}{v^4},
\end{eqnarray}
and
\begin{eqnarray}
\lambda_4^{eff}= 3 \, \left(1- 2C_{h}^{K} \right)\frac{m_h^2}{v^2}
+6\lambda_2\frac{v^2}{\mathcal{M}^2} - \frac{19N_c}{4  \, \pi^2} \, \frac{m_t^4}{v^4}.
\end{eqnarray} 
With 
$v = 246 \, {\rm GeV}$, $m_t = 174 \, {\rm GeV}$ and 
${\mathcal{M}} =1$~TeV, and taking $m_h=v/2$,  these are 
\begin{eqnarray}
\lambda_3^{eff} &=& 0.62-0.05(C^1_\phi+\tfrac14C^2_\phi)+0.06\lambda_2, \\
\lambda_4^{eff}&=& 0.39-0.09(C^1_\phi+\tfrac14C^2_\phi)+0.36\lambda_2.
\end{eqnarray} 
For negative  $\lambda_2$
 of order one one can greatly reduce the repulsive contact
 interaction, $\lambda^{eff}_4$,  in a putative higgs-higgs bound state. Of
 course, this comes at the price of reducing the attractive
 interaction, governed by $\lambda^{eff}_3$.

\subsection{$g \, g \to h \, h$  Production}

\begin{figure}[hbtp]
\centerline{\scalebox{1}{\includegraphics{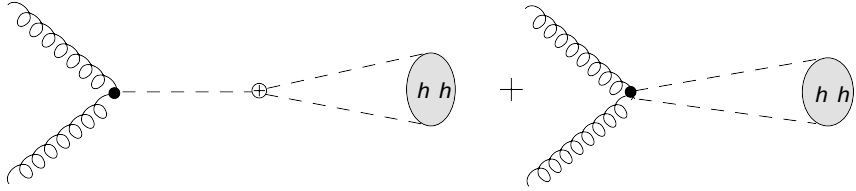}}}
\caption{The two higgs production process in the effective theory. }
\label{boundstate1}
\end{figure}

From our results in Section \ref{twohiggsprod}, the production of two higgs in our effective theory framework is
straightforward to write down. The contributions to the amplitude are
shown in Fig.~\ref{boundstate1}. 

The amplitude for two higgs production, to $\orderalpha$, is given by
\begin{eqnarray}
\langle h \, h | i \, A |  A^\alpha(P_1) \, A^\beta(P_2) \rangle  &=&\langle h \, h | i \, A_1 |  A^\alpha(P_1) \, A^\beta(P_2) \rangle 
+ \langle h \, h | i \, A_2 |  A^\alpha(P_1) \, A^\beta(P_2) \rangle
\end{eqnarray}
where we have
\begin{eqnarray}
\langle i \, A_1 \rangle^{\alpha \, \beta}  &=& 2 \, i  \, C_F  \, (C^{eff}_{G\, G \, h} )\, \frac{f^{\alpha \, \beta}(P_1,P_2)}{(P_1 + P_2)^2 - m_h^2 + i \, \epsilon} \, \left( v \, \lambda_3^{eff} - \frac{2}{v} \, C_H^K \, \left(P_3^2 + P_4^2 + (P_1 + P_2)^2 \right)\right), \nn \\
\langle i \, A_2 \rangle^{\alpha \, \beta} & =& 4 \, i  \, C_F  \, (C^{eff}_{G\, G \, h \, h} ) \, f^{\alpha \, \beta}(P_1,P_2),
\end{eqnarray}
where 
\begin{eqnarray}
f^{\alpha \, \beta}(P_1,P_2) \equiv P_1^{\alpha} \, P_2^{\beta}  + P_1^{\beta} \, P_2^{\alpha}   - 2 \, g^{\alpha \, \beta} \, P_1 \cdot P_2.
\end{eqnarray}

Using two higgs production as a test of the cubic self coupling of the
higgs has been examined in \cite{Dawson:1998py} where testing for the
MSSM with this signal was investigated. Ref.~\cite{Pierce:2006dh}
advocated the examination of $g \, g \to h \, h$ to compare the one and
two higgs production coefficients in Eq.~\eqref{hvshh} since the naive
relation between the two coefficients could be upset by the presence
of novel operators like that in \eqref{eq:pierce}.  As we have discussed, in the 
non linear realization of broken electro-weak symmetry, the relationship between $g \, g \to h \,
h$ production and $g \, g \to h $ production is not fixed as in the linear realization.
Any deviation from the SM value for $g \, g \to h \, h$ must be interpreted with care.
 The $g \, g \to h \, h$ production rate in our effective theory construction (in the linear realization) depends on at least six
unknowns, namely, ${\mathcal{M}}, \lambda_2, C_{\phi}^1, C_{\phi}^2, c_G,
\tilde{c}_G$. The effects of the operator advocated in
\cite{Pierce:2006dh} increase the number of unknown parameters
still further. 

Clearly two higgs production is an important signal to test the higgs
mechanism in the standard model.  The cross section of $g \, g \to h \,
h$ is suppressed compared to the cross section of $g \, g \to h $ by a
factor of 1000, due to the effects of parton distribution functions and phase space
suppression\cite{Dawson:1998py}. The cross section falls from $50 \,
{\rm fb} $ to 10 $\rm fb$ as the higgs ranges in mass from $100 \,
{\rm GeV}$ to $200 \, {\rm GeV}$.  Thus once LHC enters its high
luminosity running of $100 \, {\rm fb^{-1}}/{\rm Year}$ one can expect
roughly $1000$ events per year.  A significant excess or deficit of
this signal should be observable. However, the reconstruction of  exactly what
form of new physics is present requires more information.

One could obtain more information on the unknown parameters involved
by further probes of the physics of the self interaction of the higgs.
In the remainder of the paper we examine the sensitivity of a higgs
bound state (Higgsium) in the appropriate low energy effective field
theory on $\rm TeV$ scale physics to these parameters.  If a bound state forms, one can
use the properties of the bound state such as its binding energy, as
an a probe of the physics above the scale $\mathcal{M}$.

\subsection{Higgsium: Production and Decay time}

To get some rough understanding of the conditions under which a
higgs-higgs bound state may form, consider the 
non-relativistic Schrodinger equation
\begin{eqnarray}
\left[- \nabla_r^2 + V(r) - E \right] \, \psi(r) =0,
\end{eqnarray}
with the potential from a yukawa exchange and a contact interaction, 
\begin{eqnarray}
V(r) = - \frac{g^2}{4 \, \pi} \, \frac{e^{- m_h \, r}}{r}  + \kappa \, \delta^3(r).
\end{eqnarray}
We are interested in the case $g \sim  \lambda_3^{eff}$ and $\kappa \sim  \lambda^{eff}_4$ 
as a non relativistic approximation of the higgs self interactions.  
Neglect for now  the contact interaction. It is
well known that the Yukawa potential produces bound states provided
\begin{eqnarray}
 \frac{g^2}{4\pi} \gtrsim  1.7, 
\end{eqnarray} 
Neglecting new physics effects,
\begin{eqnarray}
 \lambda_3^{eff} \approx  \, 3 \, \frac{m_h^2}{v^2},
\end{eqnarray} 
 non-relativistic bound states could be expected for
\begin{eqnarray}
m_h  \gtrsim  1.2 v.
\end{eqnarray} 
The effect of TeV scale new physics changes the relationship between
the mass and the coupling. The above Yukawa bound state condition is
modified to
\begin{eqnarray}
m_h  > \left[ 1.54  +0.09(C^1_\phi+\tfrac14C^2_\phi)- 0.02 \, \lambda_2 \right]^{\!\frac12} \, v.
\end{eqnarray}  
This demonstrates the point that if the higgs self coupling is
significantly stronger due to strong TeV scale new physics that
contributes large Wilson coefficients, then a low energy signal of
this higher scale physics might be a NR bound state
formed by two higgs.

However, one can see that it is difficult to realize the NR bound state
condition when we identify the couplings in this Schrodinger equation
with our effective couplings. This identification is in fact
incorrect. We will demonstrate in Section \ref{NRHET} that the correct
NR limit of the higgs sector is described by a Lagrangian containing
only contact interactions and higher derivative operators.

The formation time of the bound state can be approximated by
the ratio of $ 4 \, R_0 / u$ where $R_0$ is the characteristic radius of the
NR bound state and  $u$ is the relative velocity of the two higgs. This is roughly the
period of oscillation for S wave states \cite{Strassler:1990nw}.

For a NR bound state we can approximate 
the relative momenta of the two higgs by $p \sim m_h \, u$ so that 
\begin{eqnarray}
\tau_f &\sim& \frac{4 \, R_0}{u}, \nn \\
&\sim& \frac{4}{m_h  u^2}. 
\end{eqnarray}  

The SM higgs decays predominantly via $h \to b \, \bar{b}$ pairs through 
Yukawa interactions if $114.4 < m_h \ll 2 \, M_{Z}$.  We take these decays as dictating
the decay width of Higgsium.

Neglecting the effects of our new operators, this decay has the decay width 
\begin{eqnarray}
\Gamma_b = \frac{m_b^2}{v^2} \, \frac{3 \, m_h}{4 \, \pi} \, \left( 1 - 4 \, \frac{m_f^2}{m_h^2} \right)^{3/2}.
\end{eqnarray}
This gives an approximate decay time
\begin{eqnarray}
\tau_b = \frac{4 \, \pi}{3 \,m_h} \, \frac{v^2}{m_b^2}.
\end{eqnarray}
The condition that the bound state has time to form is that $\tau_f < \tau_b$
which can be satisfied for 
\begin{eqnarray}
u^2 > \frac{3}{\pi} \, \left( \frac{m_b^2}{v^2} \right).
\end{eqnarray}
Thus a non relativistic bound state has time to form before it decays.
Above $135 \, {\rm GeV}$ and below the threshold of $W^+ \, W^-$
production,  the dominant decay of the higgs is through a
virtual $W$ pair,  $h \to W \, W^\star$.  Above $m_h > 2 \, m_W$ the decay into 
$W^+ \, W^-$ predominates and the decay width is given by
\begin{eqnarray}
\Gamma_W &=& \frac{m_h^3}{v^2} \, \frac{1}{32 \, \pi} \, \sqrt{1- a_W} \, \left( 4 - 4 \, a_W + 3 \, a_W^2 \right),
\end{eqnarray}
where $a_W = 4 \, m_W^2/m_h^2$ using the notation of \cite{PDBook}.
Comparing the formation and decay time for a higgs whose mass is above the threshold of $W^+ \, W^-$
production we find
\begin{eqnarray}
u^2 >   \frac{1}{\eta(m_h,m_w)} \, \frac{3}{8 \, \pi} \, \left( \frac{m_h^2}{v^2} \right),
\end{eqnarray}
where $ \eta(m_h,m_w) \sim 1$. 

The lower bounds on $u$ in either case are
compatible with the NR approximation for the full range of higgs
masses we consider.
A relativistic bound state
is also possible in either case, although an approximation scheme that
can estimate its formation time is lacking.  In the remainder of the
paper we focus on the possibility of a NR bound state being formed by
a relatively light higgs,  $m_h < 2 \, m_t$, due to our treatment of the top quark. 
We also note that a NR bound state may also have observable effects on the spectrum of two higgs
production even if a bound state does not fully form as in the 
case of top quark pair production near threshold in $e^+e^-$ collisions \cite{Strassler:1990nw}. 
 
\section{Non Relativistic higgs Effective Theory}\label{NRHET}

If the two higgs are created with small relative velocity and form a
non relativistic bound state it is appropriate to describe the physics
of this state with a non relativistic effective field theory of the
higgs sector.  We refer to our effective theory derived in Section
\ref{HET} through Section \ref{twohiggsprod} as higgs Effective Theory
(HET) and now match onto a non-relativistic version of this theory
(NRHET) where we take the $c \to \infty$ limit of the scalar field Lagrangian
density of HET.  Recall the Lagrangian density is of the form
\begin{equation}
\mathcal{L} =
 \frac{1}{2}\left[1+c_1^{eff}\frac{h}{v}+c_2^{eff}\frac{h^2}{v^2}\right]
 \, \partial^\mu \, h \,  \partial_\mu \, h - \frac{1}{2} \, m_h^2 \, h^2 - \frac{v \, \lambda^{eff}_3}{3 \, ! } \, h^3  - \frac{\lambda^{eff}_4}{4 \, ! } \, h^4 +  \mathcal{O} \left(\frac{v^2}{\mathcal{M}^2} \right).
\end{equation}
We wish to construct the non relativistic limit of this Lagrangian
density systematically, retaining $\hbar = 1$ and making factors of $c$
explicit with $[c] \sim [x]/[t]$.  The dimensionful quantities can be
expressed in units of length $[x]$ and time $[t]$.  As $\hbar = 1$, we
still have $[E] \sim 1/[t]$ and $[p] \sim 1/[x]$.  As the action $S = \int dt
\, d^3x \, \mathcal{L}$ is dimensionless, we have $[\mathcal{L}] \sim
[x]^{-3} \, [t]^{-1}$.  For the time and spatial derivatives to have
the same units in $\mathcal{L}$ we take
\begin{eqnarray}
\partial_0 = \frac{1}{c} \, \frac{\partial}{\partial \, t},
\end{eqnarray}
and so $\partial^{\mu} \sim 1/ [x]$.  This gives $[h] \sim 1/\sqrt{[x] \, [t]}$.  We
require $ [m_h \, c^2] \sim [E] \sim 1/[t]$, so that we have $[m_h] \sim
[t]/[x]^2$, and choose the electroweak symmetry breaking expectation
value to have the same dimensions as the field $h$, $[v] \sim 1/\sqrt{[x]
\, [t]}$.  The Lagrangian density with these unit conventions is given
by
\begin{eqnarray}
\mathcal{L} =
 \frac{1}{2}\left[1+c_1^{eff}\frac{h}{v}+c_2^{eff}\frac{h^2}{v^2}\right]
 \partial^\mu  h \,  \partial_\mu  h - \frac{1}{2} \, m_h^2 \,c^2\, h^2 - \frac{v \,
 \lambda^{eff}_3}{3 \, ! \, c} \, h^3  - \frac{\lambda^{eff}_4}{4 \, ! \, c} \, h^4 +  \mathcal{O} \left(\frac{v^2}{\mathcal{M}^2} \right)
\end{eqnarray}

Now consider the non-relativistic limit of this theory. The
interaction terms will be determined below by matching. Consider first
the theory of a free real scalar field of mass $m_h$ given by
\begin{equation}
\mathcal{L}=\frac12\,\partial^\mu  \varphi  \,  \partial_\mu  \varphi  - \frac{1}{2} \, m_h^2 \,c^2\, \varphi^2. 
\end{equation}
The field $\varphi$ must also be expanded in the $c \to \infty$ limit. 
We remove a large energy scale $m_h \, c^2 $ from this field 
with a field redefinition
\begin{eqnarray}
\varphi(x) =   e^{- i \,  m_h \, c^2 \, r \cdot x} \, \varphi_{+}(x) +   e^{i \,  m_h \, c^2 \, r \cdot x} \, \varphi_{-}(x) ,
\end{eqnarray}
where $r = (1,{\bf{0}})$ and $\varphi_+(x), \varphi_-(x)$ correspond
to the creation and annihilation components of the scalar field
$\varphi(x)$.  Expanding the Lagrangian density in terms of
$\varphi_\pm(x)$ we neglect terms multiplied by factors of
\begin{eqnarray}
e^{n \, i \, m_h \, c^2 t }, \quad (n\neq0, n \in {\mathcal{I}}).
\end{eqnarray}
These are terms in the Lagrangian density where some of the fields
are far off shell. Their effect is only to modify coefficients of local
operators in the effective action. 

With this substitution we find
\begin{equation}
\mathcal{L} =  \partial_0  \, \varphi_- \,  \partial^0 \, \varphi_+  
-  \partial_i  \, \varphi_- \,  \partial^i \, \varphi_+ 
 +  i \, m_h \, c\, \left(\varphi_- \,  \partial^0 \, \varphi_+ - \varphi_+ \,  \partial^0 \, \varphi_- \right) . 
\end{equation}
The first term, with two time derivatives, is suppressed by $1/c^2$
and is suppressed in the $c \to \infty$ limit.  Integrating by parts the
remaining kinetic terms and rescaling $h_\pm =\sqrt{2m_h}\,\varphi_\pm $ gives
\begin{equation}
\mathcal{L}_{\text{NR}}^0 = h_- \,  \left( i \,  \frac{ \partial}{\partial t} + \frac{\nabla^2}{2m_h} \right)  h_+ . 
\end{equation}

One can extend this effective Lagrangian by adding interactions,
including higher order terms
suppressed by $ |{\bf{u}}|/c$ and $v^2/ \mathcal{M}^2$ where
${\bf{u}}$ is the relative velocity of the two  higgs in a
non-relativistic bound state. Scattering is described by a contact
interaction which can be parameterized by a coupling $ C_{NR}$,
\begin{equation}
\label{eq:NRh}
\mathcal{L}_{\text{NR}} = h_- \,  \left( i \,  \frac{ \partial}{\partial t} +
\frac{\nabla^2}{2m_h} \right)  h_+  
+\frac{ C_{NR}}{4c}h_-^2h_+^2.
\end{equation}
There is no cubic interaction because this necessarily involves at
least one far off shell particle. The effect of the cubic interaction
in the HET is incorporated in the coupling $ C_{NR}$, and we will
compute this in terms of the parameters of the HET below, in
Sec.~\ref{match}.

In this effective theory, the energy and the momenta of
the system are given by
\begin{eqnarray}
k^0 &=& \frac{1}{2} \, m_h \,  |{\bf{u}}|^2, \\
{\bf{q}} &=& m_h \,   {\bf{u}},
\end{eqnarray}
where $ |{\bf{u}}| \ll c$ is the relative velocity of the two higgs.
It is advantageous to have power counting
rules in $ |{\bf{u}}|$ that are as manifest as possible in the
Lagrangian density as demonstrated in \cite{Luke:1996hj}.  We rescale
so that the natural sizes of the coordinates are given by the above
energy and momentum and define a new field $\rm H_\pm(x)$ and new, dimensionless
coordinates $\bf{X}$ and $T$ by
\begin{eqnarray}
{\bf{x}} = \lambda_x \, {\bf{X}},   \qquad\   t = \lambda_t \, T,   \qquad\  h_\pm(x) = \lambda_h \, H_\pm(x).
\end{eqnarray}
To ensure the rescaled energy and momenta are of order unity we have
$\lambda_t = m_h \lambda_x^2$ and
\begin{eqnarray}
\lambda_x &=& \frac{1}{m_h \, |{\bf{u}}|}, \\
\lambda_t &=& \frac{1}{m_h \, |{\bf{u}}|^2}, \\
\lambda_h &=& m_h^{3/2} \, |{\bf{u}}|^{3/2}, \\
K^0 &=& \frac{k^0}{m_h \, |{\bf{u}}|^2 }, \\
{\bf{K}} &=& \frac{{\bf{q}}}{m_h \, |{\bf{u}}| }.
\end{eqnarray}
The form of the Lagrangian density when we implement these re-scalings
and introduce an appropriately rescaled contact coupling, $
\hat C_{NR}=4 \, m_h^2  \, C_{NR}$ is given by
\begin{eqnarray}
\label{eq:NRH}
\mathcal{L}_{NRH} &=&  H_- \,  \left( i \, \partial^0 + \frac{\nabla^2}{2} \right)
H_+  + \frac{\hat C_{NR}}{16} \, \frac{|{\bf{u}}|}{c} \, \left(H_- \right)^2 \,  \left(H_+ \right)^2.
\end{eqnarray}
This form of the Lagrangian makes power counting explicit in the small parameter
$u/c$. Physical quantities, such as the energy of  bound states, can
be equally calculated from the theories in Eqs.~\eqref{eq:NRh}
or~\eqref{eq:NRH}. Which is used is a matter of convenience: the former
has familiar dimensions while the latter has explicit power counting.\footnote{
The $c \rightarrow \infty$ limit of NR effective field theories was studied in \cite{Grinstein:1997gv}.
The reader interested in bound states at threshold in NRHET would also profit from an examination 
of the treatment of bound states at threshold in NN effective field theory, reviewed in \cite{Beane:2000fx}. }

\subsection{Matching onto NRHET}
\label{match}
To determine the matching coefficient $C_{NR}$ we take the non
relativistic limit of the $\rm h \, h \to h \, h$ scattering determined
in HET.  We neglect the running from $m_t^2$ down to our matching
scale $\mu^2 = m_h^2$ in this initial study, and perform the matching at
tree level only. 

\subsubsection{Linear Realization}

\begin{figure}[hbtp]
\centerline{\scalebox{1}{\includegraphics{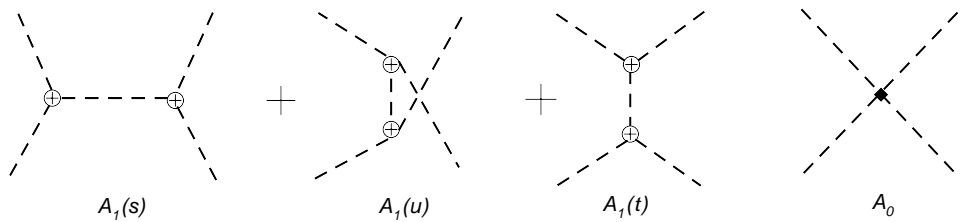}}}
\caption{Tree level $h \, h \to h \, h$ scattering in the extended higgs theory. Time flows left to right. }\label{treelevelscatt}
\end{figure}

The HET contact interaction is given by   
\begin{eqnarray}
{ \mathcal{A}}^{L}_0 =  -  3 \,  \lambda_1   + 20\,  \lambda_1  \,  C_h^K - \frac{15}{2} \, \lambda_2 \,
\frac{v^2}{\mathcal{M}^2} + \frac{4N_c}{\pi^2} \,
\left(\frac{m_t^4}{v^4}\right).
\end{eqnarray}
The Yukawa exchange feynman
diagrams, shown in Fig.~\ref{treelevelscatt}, give the amplitude
\begin{eqnarray}
i \, {\mathcal{A}}^L_y(s,t,u)  = i (A^L_1(t) +  A^L_1(u) +  A^L_1(s)) ,
\end{eqnarray}
where $s,t,u$ are  the usual Mandelstam  variables, and
\begin{eqnarray}
A^L_1(x) =  \frac{- 3 v^2 \, \lambda_1}{x - m_h^2 + i \, \epsilon} \, \left(3 \,
\lambda_1+ 5 \, \lambda_2 \, \frac{v^2}{{\mathcal{M}}} - 4 \, ( x + 2 \, m_h^2) \,
\frac{C_h^{K}}{v^2} - \frac{2N_c}{\pi^2} \,\left(\frac{m_t^4}{v^4} \right)
- 18 \, \lambda_1 \, C_h^K \right).
\end{eqnarray}

The total amplitude for $h \, h \to h \, h$ scattering  is given by
\begin{eqnarray}
{\mathcal{A}}^L_{h \, h \to h \, h}(s,t,u) =  {\mathcal{A}}^L_0 +  {\mathcal{A}}^L_y(s,t,u). 
\end{eqnarray}
To perform the matching we take the momenta of the higgs particles to
be off-shell by a small residual momenta ${\tilde{p}}$ with energy
and momenta that scale as ${\tilde{p}}_0\sim m_h u^2$ and
${\tilde{\mathbf{p}}}\sim m_h\mathbf{u}$.  The momenta of the higgs
are decomposed as (recall $r=(1,{\bf 0})$)
\begin{eqnarray}
p &=& m_h\, r + \tilde{p},  \quad \, k = m_h\, r + \tilde{k}, \nn \\
p' &=& m_h\, r + \tilde{p'}, \quad \,  k' = m_h\, r + \tilde{k'}. 
\end{eqnarray}
This gives, in the center of mass frame
\begin{eqnarray}
s &=& 4 \, m_h^2 + 4 \,  |\bf{q}|^2, \nn \\ 
t &=& - \,  |{\bf{q}}|^2 \left(1- \cos (\theta)\right), \\
u &=& - \,  |{\bf{q}}|^2 \left(1+ \cos (\theta)\right), \nn 
\end{eqnarray}
with $\mathbf{q}\sim m_h\mathbf{u}$.
In the non relativistic limit we retain the lowest order in
$|\mathbf{u}|$ and we have
\begin{eqnarray}
{\mathcal{A}}^L_{NR} &=& {\mathcal{A}}^L_0 + {\mathcal{A}}^L_1(4 \, m_h^2) + 2 \,
{\mathcal{A}}^L_1(0) \nn \\
&=&12 \,  \lambda_1+10 \, \lambda_2 \,
\frac{v^2}{\mathcal{M}^2} - 64 \, \lambda_1 \, C_h^K 
- \frac{39 \, N_c}{4 \,  \pi^2} \left(\frac{m_t^4}{v^4}\right).
\end{eqnarray}
To determine the coupling $C^4_{NR}$ in the NRHET Lagrangian,
Eq.~\eqref{eq:NRh}, we compute the four point amplitude and insist
that it equals $\mathcal{A}_{NR}$. Inserting four factors of
$\sqrt{2m_h}$ to account for relativistic normalization of states, we finally arrive at
\begin{eqnarray}
(2m_h)^2{C}^L_{NR}  =\hat{C}^{L}_{NR}
=12 \,  \lambda_1+10 \, \lambda_2 \,
\frac{v^2}{\mathcal{M}^2} - 64 \, \lambda_1 \, C_h^K 
- \frac{39 \, N_c}{4 \,  \pi^2} \left(\frac{m_t^4}{v^4}\right).
\end{eqnarray}

\subsubsection{Non-Linear Realization}
For a non-linear realization we find the following for the HET contact interaction
\begin{eqnarray}
{ \mathcal{A}}^{NL}_0 =  -\lambda_4^{eff}+4\frac{m_h^2}{v^2}c_2^{eff}.
\end{eqnarray}

The yukawa exchange diagrams give
\begin{eqnarray}
A^{NL}_1(x) =  \frac{-  v^2 }{x - m_h^2 + i \, \epsilon} \left[\lambda_3^{eff}
-\frac{c_1^{eff}}{2}\left(\frac{2m_h^2+x}{v^2}\right)\right]^2.
\end{eqnarray}
The matching is performed as in a linear realization and we find
\begin{eqnarray}
{\mathcal{A}}^{NL}_{NR} &=& {\mathcal{A}}^{NL}_0 + {\mathcal{A}}^{NL}_1(4 \, m_h^2) + 2 \,
{\mathcal{A}}^{NL}_1(0) \nn \\
&=&
\frac53\frac{v^2}{m_h^2}(\lambda_3^{eff})^2-\lambda_4^{eff}-2c_1^{eff}\lambda_3^{eff} + \left(4 \, c_2^{eff} - (c_1^{eff})^2 \right)\frac{m_h^2}{v^2}.
\end{eqnarray}

This gives the effective HET coupling  in the non-linear realization 
\begin{eqnarray}
(2m_h)^2{C}^{NL}_{NR} =\hat{C}^{NL}_{NR}= \frac53\frac{v^2}{m_h^2}(\lambda_3^{eff})^2-\lambda_4^{eff}-2c_1^{eff}\lambda_3^{eff}  
+  \left(4 \, c_2^{eff} - (c_1^{eff})^2 \right)\frac{m_h^2}{v^2}.
\end{eqnarray}

\subsection{NRHET Bound State Energy}  

\begin{figure}[hbtp]
\centerline{\scalebox{1}{\includegraphics{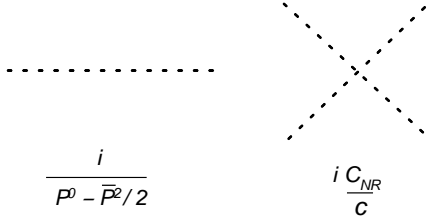}}}
\caption{Feynman rules for NRHET.}\label{NRHrules}
\end{figure}

To find the approximate bound state energy of the higgs, we calculate
the bubble sum in our NRHET theory and interpret the pole in the
re-summed bubble chain as the bound state energy of Higgsium. Note that this calculation is formally justified in the 
large $\rm N$ limit \cite{'tHooft:2002yn} where the higgs sector is equivalent to 
and $O(4)$ theory \cite{Jansen:1993jj}. The
Feynman rules for the NRHET Lagrangian in \eqref{eq:NRh} are shown in
Fig.~\ref{NRHrules}. 

The bubble sum is straightforward to calculate in NRHET. 
The leading order term is directly obtained from the Feynman rules, we use the Lagrangian given by 
Eqs.~\eqref{eq:NRh} in the following.
The leading bubble graph is given by
\begin{equation}
i \, {\mathcal{A}}_{\text{1-loop}} = \left(i \,{C}_{NR} \right)^2 \, 
\int \frac{dk^0\,d^d k}{(2  \pi)^d}\frac{i}{(E+k^0) -  {\bf{k}}^2/2m_h + i  \epsilon}  \,\cdot\,
 \frac{i}{- k^0  - {\bf{k}}^2/2m_h + i \, \epsilon}  
\end{equation}
We have chosen to work in the center of mass frame, and $E=P_1^0 + P_2^0$ stands
for the center of mass energy.  Performing the first integral by
residues and the remaining integrations with dimensional regularization, we find
\begin{equation}
\label{onebubble}
i \, {\mathcal{A}}_{1-loop} =-i \, \frac{m_h({C}_{NR})^2}{4  \pi}\, (-
m_hE)^{1/2}.
\end{equation}

\begin{figure}[hbtp]
\centerline{\scalebox{0.8}{\includegraphics{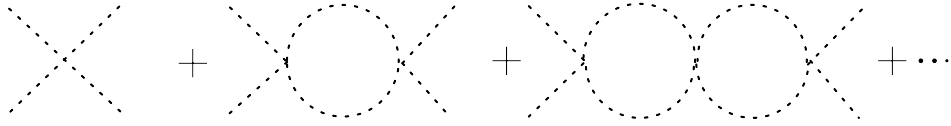}}}
\caption{The bubble sum of graphs leading to the bound state pole in
  NRHET.}
\label{diag:bubbles}
\end{figure}
The terms in the bubble sum of diagrams shown in
Fig.~\ref{diag:bubbles} are given by the geometric series
\begin{multline}
i {C}_{NR}    \left[ 1 -  \frac{m_h{C}_{NR}}{4 \, \pi}   (-m_h
  E)^{1/2} 
+ \left(  \frac{m_h{C}_{NR}}{4 \, \pi}  (-m_h E)^{1/2} \right)^2 + \cdots \right] \nn \\
=  \frac{i \,  C_{NR} }{1+  \frac{m_h{C}_{NR}}{4 \, \pi}   (-m_h E)^{1/2}}.
\end{multline}
This result agrees with \cite{Luke:1996hj,Weinberg:1991um} and
indicates a bound state with a bound state for $C_{NR}>0$ with
binding energy
\begin{equation}
E_b =  \frac1{m_h} \left(\frac{4  \pi}{m_h{C}_{NR}}\right)^2=
{m_h} \left(\frac{16  \pi}{\hat{C}_{NR}}\right)^2.
\end{equation}

There is an implicit renormalization condition introduced by dimensional
regularization. The integral has no pole as $d\to3$, so it is
interesting to ask what subtraction has been made. This is easily
understood by performing the $d=3$ integration with a momentum cut-off
$|\mathbf{k}|<\Lambda $ in terms of the bare coupling ${C}_{NR}^0$:
\begin{equation}
i \, {\mathcal{A}}_{1-loop}^\Lambda 
=i  m_h({C}_{NR}^0)^2\left[\frac{\Lambda}{2\pi^2}-\frac{1}{4  \pi}\, (-m_hE)^{1/2}\right].
\end{equation}
The renormalized coupling $C_{NR}(\mu)$ can be defined as the
 amplitude at a fixed energy $E= - \mu$ \cite{Mehen:1998zz}. Then the combination
\begin{equation}
\frac{1}{C_{NR}}\equiv \frac1{C_{NR}^0}-\frac{m_h\Lambda}{2\pi^2}=
 \frac1{C_{NR}(\mu)}-m_h\frac{(m_h\mu)^{1/2}}{4\pi}
\end{equation}
is renormalization group invariant. This is precisely the coupling
that appears in \eqref{onebubble}.

It would appear that for any positive value of $C_{NR}$ we have
bound states. However for our NR description to be self consistent we require that
the binding energy of the bound state satisfy $E_b < m_h$, that is, 
\begin{eqnarray}
\label{Cbound}
\hat C_{NR}>16\pi.
\end{eqnarray}

\subsubsection{Linear Realization}

In the case of a linear realization a heavy higgs seems necessary for
the bound state to form, but the new physics effects may allow
significantly smaller masses for the bound state.  If we neglect the
effects of new physics ($\lambda_2$ and $C_h^K$) the bound \eqref{Cbound}  translates into $m_h > 2.0 \,  v$. 
Retaining the effectsof $\lambda_2$ and $C_h^K$ one determines a condition for the NRHET
calculation of the bound state energy to be self consistent
\begin{eqnarray}
\frac{m_h}v>\sqrt{\frac{{16\pi}-4 \, \lambda_2\frac{v^2}{\mathcal{M}^2}+\frac{51 \, N_c}{4 \, \pi^2}\left(\frac{m_t^4}{v^4}\right)}{12-40 \, C_h^K}}.
\end{eqnarray}

Alternatively, for a given value of the higgs mass, say $m_h=\xi \,  v$, the
self-consistency condition implies a constraint on the coefficients of the
higher dimension operators:
\begin{equation}
12 \, \xi^2 + 4 \, \frac{v^2}{\mathcal{M}^2} \left( \lambda_1 - 10 \, \xi^2 \, (C_\phi^1+\tfrac14C_\phi^2) \right) > 16 \, \pi + \frac{51 \, N_c}{4 \, \pi^2}\left(\frac{m_t^4}{v^4}\right)
\end{equation}
Using $\mathcal{M}=1$~TeV and the PDG value for the top quark mass, this condition 
simplifies to
\begin{equation}
1.2 \, \xi^2 + 0.024 \lambda_1 - 0.24 \, \xi^2 \, (C_\phi^1+\tfrac14C_\phi^2)  > 5.1
\end{equation}
So, for example, for $|C_\phi^1+\tfrac14C_\phi^2|=1$ or 5 the minimal higgs
mass for a NR bound state is reduced by 6\% or 28\%, respectively. Near the  
limit of validity of our calculation $m_h \sim 2 \, m_t$, for negative values of 
$C_\phi^1+\tfrac14C_\phi^2$ we find that a bound state is possible for 
${\mathcal{O}}(1)$ wilson coefficients as we illustrate in Fig 7.
\begin{figure}[htb]
\centerline{\scalebox{1.1}{\includegraphics{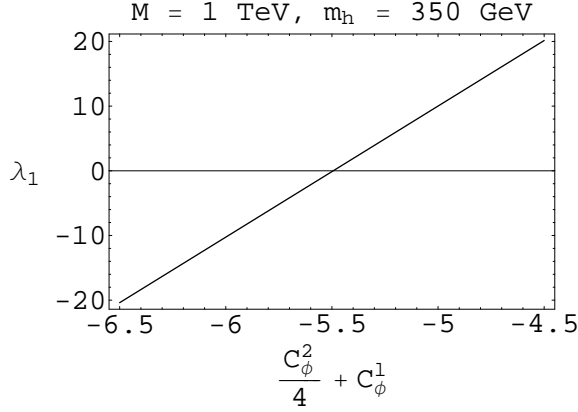}}}
\caption{In the linear realization the allowed 
parameter space for NR bound state formation is above the line.}
 \end{figure}

\newpage
\subsubsection{Non-Linear Realization}

In the nonlinear realization this condition is easily satisfied even
for a light higgs, $m_h<v$.  Recall that $\lambda_3^{eff}$ and $c_1^{eff}$
are both enhanced by powers of $\mathcal{M}/v$. Taking, for example, $m_h=120 \, {\rm GeV}$ and ${\mathcal{M}}= 1 \, {\rm TeV}$, neglecting the contribution of 
$c_2^{eff}$, the NR bound state condition is
\begin{equation}
\frac53 \, \frac{{\cal{M}}^2}{m_h^2} \,(\tilde{\lambda_3}^{eff})^2 - 2 \,\tilde{c_1}^{eff} \, \tilde{\lambda_3}^{eff} - \frac{m_h^2}{{\cal{M}}^2} \, (\tilde{c_1}^{eff})^2 >16 \, \pi + \lambda_4,
\end{equation}
where
\begin{eqnarray}
\lambda_3^{eff} &=& \left( \frac{{\mathcal{M}}}{v} \right) {\tilde{\lambda}_3}^{eff}, \nn \\
c_1^{eff} &=& \left( \frac{v}{{\mathcal{M}}} \right) \,{\tilde{c}_1}^{eff}.
\end{eqnarray}
\begin{figure}[htb]
\centerline{\scalebox{1.1}{\includegraphics{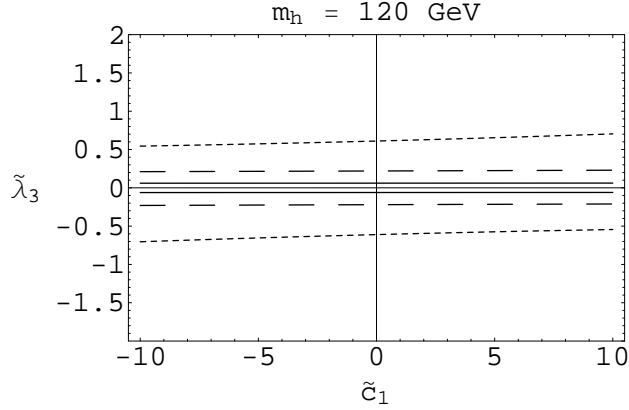}}}
\caption{In the Non-linear realization, holding $m_h$ fixed and set $\lambda_4 = 0$  as it is ${\mathcal{O}}(1)$ and suppressed by $16 \, \pi$. We vary ${\mathcal{M}}$ for the values 
 ${\mathcal{M}} = 1 \, {\rm TeV}$ (dotted line), ${\mathcal{M}} = 3 \, {\rm TeV}$ (dashed line) and  ${\mathcal{M}} = 10 \,  {\rm TeV}$ (solid line). 
 The region above (the upper) and below (the lower) hyperbolic curves satisfy NR the bound state condition. }
 \end{figure}
 \begin{figure}
\centerline{\scalebox{1.1}{\includegraphics{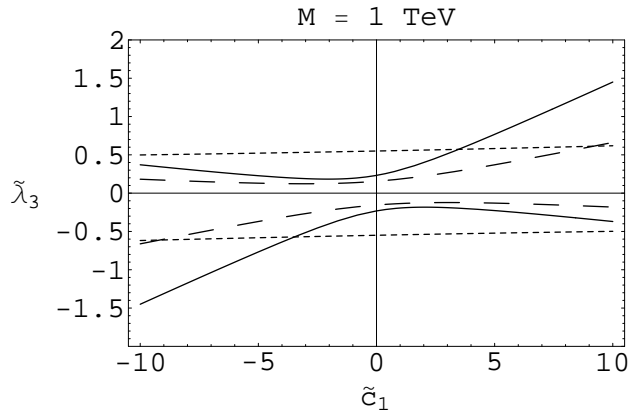}}}
\caption{In the Non-linear realization, holding ${\mathcal{M}}$ fixed and set $\lambda_4 = 0$  as it is ${\mathcal{O}}(1)$ and suppressed by $16 \, \pi$. 
We vary  $m_h = 300 \, {\rm GeV}$ (solid line), $m_h = 200 \, {\rm GeV}$ (dashed line) and  $m_h = 100 \,  {\rm GeV}$ (dotted line).
 The region above  (the upper) and below (the lower) hyperbolic curves satisfy the NR bound state condition. }
 \end{figure}

Note that as ${\mathcal{M}}$ grows larger the region that satisfies the NR bound state condition {\it grows}.
This is due to the fact that the attractive interaction given by  $\lambda_3^{eff} \sim \mathcal{M}/v$ is a relevant operator.  
We find that as  $m_h$ grows and as $\mathcal{M}$ is larger the 
allowed parameter space of the NR bound state condition is significant, demonstrating 
that a bound state is likely to form in the non linear realization.

\newpage
\section{Summary and Conclusions}
If a new strong interaction is responsible for electroweak symmetry
breaking but a higgs particle, the pseudo-goldstone boson of broken
scale invariance, remains unnaturally light, the self-interactions of
this higgs particles could be quite strong. If strong enough these
self-interactions could bind two higgs particles.

To study these questions we formulated two different effective theories
of the light, self-interacting higgs below the scale $\mathcal{M}$ of
the new physics.  In the first, the symmetry is realized linearly and
the higgs field is described as one component of an ${\rm SU}(2)_L$ doublet,
just as in the standard model of electroweak interactions. In the
second approach the symmetry is realized non-linearly: the triplet of
would-be goldstone bosons and the higgs field are not in a common
multiplet. We note that operators of dimension 3 in the effective
Lagrangian in the non-linear realization are naturally expected to be
enhanced by a power of $\mathcal{M}/v$ relative to their linear
realization counterparts. 

In order to study how large these
couplings need be, we have studied the
case of non-relativistic bound states. To this end we constructed a
non-relativistic higgs effective theory (NRHET) describing self-interacting higgs
particles in the rest frame of the bound state, in the non-relativistic limit.

The effects of the top quark are small but non-negligible. We
estimated them by including the virtual top quark effects as a
modification to the couplings in the NRHET.

Our results show, perhaps not surprisingly, that in the non-linear
realization it is quite easy to form light Higgsium, as we call the higgs-higgs bound state. 
For natural couplings in the linear realization a bound state is only likely to form for $m_h \sim v$.  
Relativistic bound states are possible in both the linear and nonlinear realizations. 

There are many questions that we have not addressed. The most
immediate one is how to search for Higgsium. Assuming a light higgs is
found, one could imagine strategies involving invariant mass
distributions of higgs-pair production. A dedicated study is required
to determine if this or other strategies are viable. Another, related
question is whether  the effects of a short lived bound state could be
seen indirectly, much like  would-be toponium affecting the line shape
in top quark pair production near threshold in $e^+e^-$ collisions. It
would also be interesting to solve the bound
state equation in the more general, fully relativistic case. We hope
to return to these problems in the future.

\begin{acknowledgments}
Work supported in part by the US  Department of Energy
under contract DE-FG03-97ER40546.
\end{acknowledgments}

\appendix
\section{Custodial Symmetry and the S parameter}
There is some confusion in the literature regarding custodial symmetry and the operator 
\begin{eqnarray}
- \frac{c_{W \, B} \, g_1 \, g_2}{{\mathcal{M}}^2} \, \left(\phi^\dagger \, \sigma^I \, \phi \right) \, B^{\mu \, \nu} \, W_{I \, \mu \, \nu} 
\end{eqnarray} 
which corresponds to the S parameter. Consider the matrix representation of this operator \cite{Willenbrock:2004hu} where the Higgs doublet field is given by
\begin{eqnarray}
\phi = \left(
 \begin{array}{c} 
 \phi^{+} \\
\phi^0 
\end{array}
\right).
\end{eqnarray}
Then $\epsilon \, \phi^\star$ is also an $\rm SU_{L}(2)$ doublet with components
\begin{eqnarray}
\epsilon \, \phi^\star = \left(
 \begin{array}{c} 
 \phi^{0 \, \star} \\
- \phi^- 
\end{array}
\right),
\end{eqnarray}
where $\phi^- = \phi^{+ \, \star}$. The Higgs bi-doublet field is given by
\begin{eqnarray}
\Phi &=& \frac{1}{\sqrt{2}} \, \left(\epsilon \, \phi^\star, \phi \right), \nn \\
&=& \frac{1}{\sqrt{2}} \,  \left(
 \begin{array}{cc} 
 \phi^{0 \, \star} &  \phi^{+} \\
- \phi^-  & \phi^0
\end{array}
\right).
\end{eqnarray}
The $\rm SU_L(2) \times U_{Y}(1)$ gauge symmetry acts on the Higgs bi-doublet as
\begin{eqnarray}
SU_L(2)&:&  \Phi \rightarrow L \, \phi \\
U_{Y}(1)&:&  \Phi \rightarrow \Phi \, e^{- i \sigma_3 \, \theta/2}. 
\end{eqnarray}
In the limit that hyper charge vanishes the Lagrangian also has the following global symmetry
\begin{eqnarray}
SU_R(2)&:&  \Phi \rightarrow \phi \, R^\dagger.
\end{eqnarray}
When the Higgs acquires a vacuum expectation value, both $\rm SU_L(2)$ and $\rm SU_R(2)$ are broken, however the 
subgroup $\rm SU_{L=R}(2)$ is unbroken, ie 
\begin{eqnarray}
L \, \langle \Phi \rangle \, L^\dagger =  \langle \Phi \rangle.
\end{eqnarray}
This is explicitly the custodial symmetry, and the corresponding transformation of the Higgs bi-doublet under this symmetry.
It is easy to see that  
\begin{eqnarray}
- \frac{c_{W \, B} \, g_1 \, g_2}{{\mathcal{M}}^2} \, {\rm Tr} \left(\Phi^\dagger \, \sigma^I \, W_{I \, \mu \, \nu}  \Phi \right) \, B^{\mu \, \nu}  
\end{eqnarray} 
is invariant under this symmetry. The Higgs bi-doublet transforms as above and the field strength $\sigma^I \, W_{I \, \mu \, \nu}$ transforms as
\begin{eqnarray}
\sigma^I \, W_{I \, \mu \, \nu}   \rightarrow L  \, \sigma^I \, W_{I \, \mu \, \nu} \, L^\dagger.
\end{eqnarray} 
However, it is also easy to see that this representation of the operator vanishes by explicitly performing the trace; one finds
\begin{eqnarray}
 {\rm Tr} \left(\Phi^\dagger \, \sigma^I \, W_{I \, \mu \, \nu} \, \Phi \right) = 0.
\end{eqnarray} 
The non trivial representation of the operator in terms of the bi-doublet is given by
\begin{eqnarray}
-{\rm Tr} \left(\Phi^\dagger \, \sigma^I \, \Phi  \, \sigma^3 \right). 
\end{eqnarray} 
With this factor of $\sigma^3$, required for a non trivial representation in terms of the Higgs bi-doublet, one finds that this operator violates custodial symmetry.

\section{Top Quark OPE} \label{opeappend}
As an example of the effect of the neglected terms in in the top quark OPE, consider the 
OPE corrections to the four point function of the higgs. The amplitude is given by
\begin{eqnarray}
i \, A_4(s,t,u) &=& - 6 \, N_C  \, \left(\frac{ m_t}{v}\right)^4 \int \frac{d^d \, k}{(2 \pi)^d} {\rm Tr}[ \frac{\left( k \! \! \! / + m_t \right)}{k^2 - m_t^2}   \frac{\left( k \! \! \! / +  a \! \! \! /  + m_t \right)}{(k+a)^2 - m_t^2}
 \frac{\left( k \! \! \! / +  b \! \! \! /  + m_t \right)}{(k+b)^2 - m_t^2}
 \frac{\left( k \! \! \! / +  c \! \! /  + m_t \right)}{(k+c)^2 - m_t^2}]. \nn
\end{eqnarray}
We find the leading order in $p^2/m_t^2 \to 0$ the amplitude is given by
\begin{eqnarray}
i \,  A_4^0(s,t,u) &=&  - 24 \, N_C \,  \left(\frac{m_t}{v}\right)^4 \int \frac{d^d \, k}{(2 \pi)^d} \, \frac{ \left( m_t^4 + 6 \, k^2 \, m_t^2 + k^4\right)}{\left( k^2 - m_t^2 \right)^4}, \nn \\
 &=& - \frac{i \, N_c}{16 \, \pi^2} \, \left(\frac{m_t}{v}\right)^4 \left(\frac{24}{\epsilon}  - 64 + 24 \, \log \left[ \frac{\mu^2}{m_t^2} \right] \right).
\end{eqnarray}
The  leading order matching gives a factor of $- 4 \, N_C \, m_t^4/v^4$.

Consider performing the top quark OPE to higher orders.
We find  for the next order in $p^2/m_t^2$
\begin{eqnarray}
i \, A_4^1(s,t,u)&=& - \frac{i \, N_C}{16 \, \pi^2} \, \left(\frac{m_t}{v}\right)^4 \, \left( \frac{1}{80 \, m_t^2} \right) \, \left(a^2 + b^2 + c^2  +a \cdot b+ 6 \, a \cdot c + b \cdot c \right).
\end{eqnarray}
The invariants of the external momenta $a,b,c$ averaged over the sum of all $A_4$ diagrams can be expressed in the Mandelstam variables.
We find that  our momenta expressed in terms of these variables are
\begin{eqnarray}
\langle  a^2 \rangle & = & 4 \, ! \, m_h^2, \nn \\
\langle  b^2 \rangle & = & 8 \left(s + t +u \right), \nn \\
\langle  c^2 \rangle & = & 4 \, ! \, \left[ \left(s + t +u \right) - 3 \, m_h^2 \right], \nn \\
\langle  a \cdot b \rangle & = & 4 \, \left(s + t +u \right), \nn \\
\langle  a \cdot c \rangle & = & 8 \, \left(s + t +u - 3 \, m_h^2 \right), \nn \\
\langle  b \cdot c \rangle & = & 16 \, \left(s + t +u - 3 \, m_h^2 \right).
\end{eqnarray}

With these substitutions, the next order in the expansion gives
\begin{eqnarray}
i \, A_4^1(s,t,u)&=& -  \frac{i \, N_C}{16\,  \pi^2} \,  \left(\frac{m_t}{v}\right)^4 \, \left(\frac{m_h^2}{m_t^2}\right)  \, 
\left( \frac{s + t + u}{4 \, m_h^2}  - \frac{3}{5} \right), \nn \\
&=&  - \frac{i \, N_C }{16 \,  \pi^2} \,  \left(\frac{m_t}{v}\right)^4 \, \left(\frac{m_h^2}{m_t^2}\right) \, \frac{2}{5}.
\end{eqnarray}
Where in the last expression we simplified with $s+ t + u = 4 \, m_h^2$.  
This term matches onto the operator  
\begin{eqnarray}
O^{2,0}_h = \frac{h \, h}{{\mathcal{M}}^2} \, \partial^\mu \, h \, \partial_\mu \, h,
\end{eqnarray}
with a Wilson coefficient that contains contributions from the integrating out $\rm TeV$ scale new physics and 
the top quark. At the scale $\mu^2 = m_t^2$ the Wilson coefficient is
\begin{eqnarray}
C^{2,0}_h(m_t^2)  = \frac{{\mathcal{M}}^2}{v^2} \left(4 \, C_h^K \, (m_t^2)   +  \frac{m_t^2}{v^2} \, \, \frac{N_C }{20 \, \pi^2} \right).
\end{eqnarray}
The later term in the Wilson Coefficient is an example of a term that is neglected in our calculation.
Corrections of this form can be  systematically included by taking the top quark OPE to next order in $p^2/m_t^2$.


\bibliographystyle{h-physrev3.bst}
\bibliography{higgs}
\end{document}